\pdfoutput=1
\documentclass[a4paper,11pt]{article}

\usepackage{jcappub}
\usepackage[section]{placeins}
\usepackage[cp1250]{inputenc}
\usepackage{graphicx}
\usepackage{epstopdf}
\usepackage{amsmath,amsthm,latexsym,amssymb,amsfonts,epsfig}
\usepackage{fontenc,layout}
\usepackage{subfig,graphicx}
\usepackage[usenames,dvipsnames]{xcolor}
\usepackage{verbatim}
\usepackage{relsize}
\usepackage{soul}

\newcommand{\be}{\begin{equation}} 
\newcommand{\ee}{\end{equation}}
\newcommand{\bea}{\begin{equation}\begin{aligned}} 
\newcommand{\eea}{\end{aligned}\end{equation}}

\newcommand{\td}{{\rm d}} 

\DeclareMathOperator{\erfc}{erfc}

\numberwithin{equation}{section}

\begin{document}


\title{Detecting circular polarisation in the stochastic gravitational-wave background from a first-order cosmological phase transition}

\author[1,2,3]{John~Ellis,}
\author[1]{Malcolm~Fairbairn,}
\author[1,4]{Marek~Lewicki,}
\author[1,3]{Ville~Vaskonen,}
\author[1]{Alastair~Wickens}
\affiliation[1]{Department of Physics, King's College London, Strand, WC2R 2LS London, UK}
\affiliation[2]{Theoretical Physics Department, CERN, Geneva, Switzerland}
\affiliation[3]{National Institute of Chemical Physics \& Biophysics, R\"avala 10, 10143 Tallinn, Estonia}
\affiliation[4]{Faculty of Physics, University of Warsaw ul.\ Pasteura 5, 02-093 Warsaw, Poland}
\emailAdd{john.ellis@cern.ch}
\emailAdd{malcolm.fairbairn@kcl.ac.uk}
\emailAdd{marek.lewicki@kcl.ac.uk}
\emailAdd{ville.vaskonen@kcl.ac.uk}
\emailAdd{alastair.wickens@kcl.ac.uk}

\abstract{We discuss the observability of circular polarisation of the stochastic gravitational-wave background (SGWB) generated by helical turbulence following a first-order cosmological phase transition, using a model that incorporates the effects of both direct and inverse energy cascades. We explore the strength of the gravitational-wave signal and the dependence of its polarisation on the helicity fraction, $\zeta_*$, the strength of the transition, $\alpha$, the bubble size, $R_*$, and the temperature, $T_*$, at which the transition finishes. We calculate the prospective signal-to-noise ratios of the SGWB strength and polarisation signals in the LISA experiment, exploring the parameter space in a way that is minimally sensitive to the underlying particle physics model. We find that discovery of SGWB polarisation is generally more challenging than measuring the total SGWB signal, but would be possible for appropriately strong transitions with large bubble sizes and a substantial polarisation fraction.
\\
\\
\\
\\
KCL-PH-TH/2020-25, CERN-TH-2020-070
}

\maketitle

\section{Introduction}

One of the most interesting scientific targets for upcoming gravitational-wave (GW) detectors is the stochastic gravitational-wave background (SGWB)~\cite{Allen:1996vm,Caprini:2018mtu}. This SGWB could come from many sources including primordial inflation~\cite{Bartolo:2016ami} astrophysical sources~\cite{Regimbau:2011rp} and strong phase transitions in the early Universe~\cite{Witten:1984rs,Turner:1987bw,Caprini:2009yp}.
 
The most easily measurable characteristic of a SGWB is its frequency spectrum, but this provides limited insight into its origin. Further valuable information could be provided by its intrinsic circular polarisation, which is due to a difference between the amplitudes of GWs with left and right polarisations. A SGWB generated by astrophysical sources would have negligible net polarisation, since it arises from multiple uncorrelated sources. However, a cosmological SGWB could be generated coherently over large scales, and might exhibit net circular polarisation if interactions that violate parity were important in the early Universe. Indeed, polarisation of the SGWB could in principle arise from a variety of physical mechanisms in the early Universe, for example gravitational chirality and modifications of gravity at high energies~\cite{Contaldi:2008yz,Bartolo:2017szm,Alexander:2009tp}, pseudoscalar-like couplings between the inflaton and gauge fields~\cite{Barnaby:2011qe,Adshead:2013nka,Dimastrogiovanni:2016fuu} and helical turbulence created during a first-order phase transition~\cite{Kahniashvili:2005qi, Kisslinger:2015hua}. Thus the polarisation of the SGWB could be an important diagnostic tool for probing fundamental physical processes in the early Universe.

With regards to the detectability of {polarisation of GWs}, we recall that a single linear interferometric detector cannot probe  circular polarisation in the SGWB, because it cannot distinguish between left- and right-handed GWs with the same wave vector $\vec k$. Nor can a planar  interferometer detect circular polarisation in an isotropic SGWB: it cannot  distinguish a left-handed GW with a wave vector $\vec k$ from either a right-handed GW of the same amplitude or a wave vector $\vec k'$ that is the reflection of $\vec k$ in the plane of the interferometer. However, there is a considerable body of work on the detection of polarisation in the SGWB using the orbital motion of one or multiple planar GW detectors~\cite{Seto:2006dz,Seto:2006hf,Seto:2007tn,Seto:2008sr,Romano:2016dpx,Smith:2016jqs}, and we note that initial constraints on the polarisation of the SGWB have already been provided by LIGO~\cite{Crowder:2012ik}.

As pointed out in Ref.~\cite{Seto:2006hf}, the isotropy of the SGWB is broken by our motion relative to the cosmological reference frame, which induces a dipole in the SGWB. This can be detected as a difference in the amplitudes of the GWs arriving from the $\vec k$ direction compared to those from the $\vec k'$ direction, and therefore enables a planar interferometer such as LISA to probe the possible net circular polarisation of a SGWB. This effect was used to put a lower limit~\cite{Domcke:2019zls} on the strength of a fully-polarised constant SGWB, $\Omega_{\rm GW}h^2 \simeq 10^{-11}$, in order for it to be detectable by the LISA experiment~\cite{Caprini:2015zlo} or the Einstein Telescope (ET)~\cite{Sathyaprakash:2012jk}.

In this work we focus on the polarisation of the SGWB that might have been generated by helical turbulence following a first-order phase transition in the early Universe. The Standard Model (SM) of particle physics would not have caused a first-order transition, proceeding instead via a crossover~\cite{Kajantie:1996mn}, but many proposed extensions of the SM would have. {Examples where the SGWB from a phase transition has been studied include scenarios for baryogenesis at the electroweak scale~\cite{No:2011fi,Huang:2016odd,Chala:2016ykx,Katz:2016adq,Vaskonen:2016yiu,Dorsch:2016nrg,Artymowski:2016tme,Beniwal:2017eik,Huang:2017kzu,Beniwal:2018hyi,Bruggisser:2018mrt, Huang:2018aja, Ellis:2019flb}, models with hidden sectors~\cite{Espinosa:2008kw,Schwaller:2015tja,Jaeckel:2016jlh,Huang:2017laj,Tsumura:2017knk, Baldes:2018emh, Croon:2018kqn,Breitbach:2018ddu,Croon:2018erz,Fairbairn:2019xog,Helmboldt:2019pan}, an extended gauge group~\cite{Huber:2015znp,Jinno:2016knw,Iso:2017uuu,Demidov:2017lzf,Hashino:2018zsi,Marzo:2018nov,Miura:2018dsy,Azatov:2019png} or higher-dimensional interactions among SM fields~\cite{Ellis:2018mja,Ellis:2019oqb}, and many other models~\cite{Dorsch:2014qpa,Kakizaki:2015wua,Hashino:2016xoj,Huang:2016cjm,Kubo:2016kpb,Balazs:2016tbi,Baldes:2017rcu,Marzola:2017jzl,Kang:2017mkl,Huang:2017rzf,Chala:2018ari,Megias:2018sxv,Prokopec:2018tnq,Brdar:2018num,Chala:2018opy,Baratella:2018pxi,Angelescu:2018dkk,Alves:2018jsw,Dev:2019njv,Jinno:2019jhi,Wang:2019pet,DelleRose:2019pgi,vonHarling:2019gme,Wang:2020jrd}.}
These examples illustrate how the SGWB may be an interesting probe of many scenarios for possible new fundamental physics beyond the SM, and a partnership between GW detectors, collider and other laboratory experiments could help distinguish between them.

The subject of helical MHD turbulence in the primordial plasma from such a transition has been extensively discussed in the literature~\cite{Sigl:1996dm,Kahniashvili:2009qi,Kahniashvili:2012uj,Ellis:2019tjf} and is  often looked at in the context of its impact on any potential primordial magnetic field~\cite{Kahniashvili:2012vt,Kahniashvili:2015msa,Ellis:2019tjf}. However, more recently some effort has been put into understanding the potential effects it could have on the period of GW generation expected after a first-order phase transition and whether this could be imprinted on the GW spectrum~\cite{Nicolis:2003tg,Caprini:2006jb,Kahniashvili:2009mf,Caprini:2009yp,Kisslinger:2015hua,Niksa:2018ofa,Pol:2019yex}. For the purpose of our work we are particularly interested in assessing the detectability of a potential net circular polarisation of the SGWB arising from helical MHD turbulence following a first-order phase transition. 

In order to be relatively insensitive to the details of models, we characterise them by the helicity fraction, $\zeta_*$, the strength of the transition, $\alpha$, the average bubble size, $R_*$, and the temperature, $T_*$, at which the transition is completed. We calculate the prospective signal-to-noise ratios (SNRs) of the SGWB 
frequency spectrum and polarisation signals for LISA.

The outline of our paper is as follows. In Section~\ref{sec:helturb} we review and discuss how helical turbulence in the primordial plasma may source polarisation in the SGWB generated during a first-order phase transition. In Section~\ref{Spectrum_of_SGWB} we calculate the spectrum of the SGWB, and in Section~\ref{sec:polarisation} we present our calculation of the possible polarisation of the SGWB. Section~\ref{sec:results} presents our results for the observability by LISA of polarisation in the SGWB as a function of the strength of the first-order phase transition, the bubble size and the temperature at which the tranistion occurs, as well as the assumed initial helicity fraction. Finally, Section~\ref{sec:conx} presents our conclusions.


\section{Stochastic gravitational wave background from helical turbulence}
\label{sec:helturb}

In this section we discuss relevant aspects of the epoch following the phase transition, during which we expect magnetohydrodynamic (MHD) turbulence to be generated in the primordial plasma. 

The turbulent regime develops over time due to a system of eddy currents that are first generated at the scale of the bubble radius, $R_*$, and subsequently extend over a range of both larger and smaller length scales. This network of eddies allows for plasma and magnetic energy, both initially concentrated at {the scale} $R_*$, to be spread throughout the MHD system in a `cascade' of energy, as discussed in~\cite{Ellis:2020awk}. Once MHD turbulence has fully developed at a given scale it decays freely, and we expect equipartition between the plasma and magnetic energy densities, $\rho_{M} \sim \rho_K \sim \rho_{\rm eq}$.

As we discuss further in Section~\ref{sec:inverse}, the behaviour of the turbulence can be dramatically changed if the initial magnetic field left over after the phase transition has a non-zero helical component~\cite{Cornwall:1997ms}. Such magnetic helicity could be generated via a variety of mechanisms including bubble collisions at the electroweak~\cite{Stevens_2008,Stevens_2009} or QCD~\cite{Forbes:2000gr,Kisslinger_2003} phase transitions, baryon-number-violating processes such as decaying non-perturbative field configurations, e.g. electroweak sphalerons~\cite{Copi_2008}, or even via inflation~\cite{Campanelli:2008kh}. We do not discuss further the possible origin of this helical turbulence, but parametrise it by the initial helicity fraction of the magnetic field left over after the transition.

\subsection{Direct cascade turbulence}
\label{sec:direct}

Collisions of bubbles at the end of the phase transition cause stirring of the primordial plasma on scales close to the average radius $R_*$ of the bubbles. In order to compute the characteristic velocity of such turbulent motions at the beginning of the turbulent period, we first identify two distinct forms in which the liberated vacuum energy, typically quantified by $\alpha=\rho_{\rm vac}/\rho_{\rm rad}$, is initially deposited. We express these forms quantitatively via GW efficiency factors $\kappa_i$.

First, a fraction of the available vacuum energy goes into accelerating the bubble wall, which is expressed using the GW efficiency factor $\kappa_{\rm col}$. This fraction should generally be subtracted from the total energy subsequently deposited into the plasma. However, we will be dealing with transitions that are not strong enough to produce a runaway scenario, meaning that bubble walls will reach a terminal velocity due to their friction with the surrounding plasma long before collision. Therefore, it is valid to assume that essentially all the energy is transferred to the plasma, so that $\alpha_{\mathrm{eff}}=\alpha\left(1-\kappa_{\mathrm{col}}\right)\approx \alpha$~\cite{Ellis:2019oqb}.

The vacuum energy deposited into the plasma can either be transferred into bulk fluid motion that sources GWs or can be used in heating up the plasma itself. Thus, as is common in the literature, we finally express the  fraction of vacuum energy in our GW source as the fraction transferred into fluid motion~\cite{Espinosa:2010hh,Caprini:2015zlo,Ellis:2019oqb}:
\be \label{eq:kappa_sw}
    \kappa_{\mathrm{sw}}=\frac{\alpha}{0.73+0.083 \sqrt{\alpha}+\alpha} \,,
\ee
where we have assumed for simplicity that the speed of expansion of the walls is relatively fast, with $v_w\approx 1$. We can then express the RMS fluid velocity as~\cite{Hindmarsh:2015qta,Ellis:2020awk}
\be \label{eq:U_f}
    U_f = \sqrt{\frac{3}{4} \frac{\alpha}{1+\alpha} \kappa_{\mathrm{sw}}} \,.
\ee
Assuming that all the energy left in the bulk fluid motion when the flow becomes nonlinear is converted into vortical turbulent motions of the plasma, we take $\kappa_{\rm turb} \approx \kappa_{\rm sw}$~\cite{Ellis:2019oqb} and assume that the characteristic velocity of the plasma at the beginning of the turbulent period is as shown in Eq.~\eqref{eq:U_f}.

This initial vortical fluid motion drives the formation of a hierarchy of eddy currents, first on scales around the bubble radius and subsequently on scales $\lambda \leq R_*$. This is known as the `direct energy cascade'. We parameterise the turbulent system with the quantity $\xi_M(t)$, known as the magnetic correlation length which describes the maximum scale at which the magnetic field is correlated and thus physically represents the size of the largest magnetic eddy. During the direct cascade period energy is transferred from the initial correlation scale of the turbulence $\xi_M(t_*) \simeq R_*$, to increasingly smaller scales until it is viscously dissipated as heat at the dissipation scale of the plasma, $\lambda_d$.

The distribution of magnetic and plasma energy at different scales in a direct cascade is known to follow a Kolmogorov decay law $\rho_{*,i}(\lambda,t_*) \approx \rho_{*,i}\lambda^{-2/3}$ where $i = K, M$ for kinetic and magnetic components, respectively~\cite{Kahniashvili:2008pe,Kahniashvili:2009mf}. We assume that this direct cascade period of turbulence lasts for a few times longer than the characteristic turn-over time of the largest eddy, $\tau_0=R_*/U_f$, so that the hierarchy of eddy currents have time to equilibrate.\footnote{The eddy turn-over time is defined as the time it takes for an eddy at a given plasma scale to complete one full revolution.} Thus the duration of this stage of turbulence is $\tau_{\rm direct} = s_0\tau_0$, where in this paper we {make the representative choice} $s_0 = 3$~\cite{Kahniashvili:2008pe}.

\subsection{Inverse cascade turbulence}
\label{sec:inverse}

We expect helicity to be conserved in a highly conductive plasma. Thus, if there is some initial helicity left over in the magnetic field after the phase transition, which we parameterise with the initial magnetic helicity fraction $\zeta_*$ as defined in~\cite{Kahniashvili:2008pe}, we expect it to be approximately conserved during the direct cascade period of turbulence. After this stage the turbulence, with the plasma and the magnetic field both in equipartition, relaxes to a fully helical state, since the non-helical turbulent energy is fully dissipated away at small scales in contrast to the conserved helical component~\cite{Ellis:2020awk}.

This results in a second period of `inverse cascade' turbulence following the direct cascade stage, during which the remaining fully helical turbulence can only be transferred to scales that are increasingly larger than the bubble radius. For large enough initial helicity fractions, this can result in a rapid increase in the magnetic correlation length of the turbulence, $\xi_M(t)$, which corresponds physically to a large increase in the size of the largest eddy, compared with the direct cascade period.

Adopting Model B outlined in detail in~\cite{Kahniashvili:2008pe} and originally based on the work of~\cite{Banerjee03,Campanelli07}, we take the following evolution law for the magnetic eddy correlation scale
\begin{equation} \label{Eq:correlation-scale}
    \xi_M(t) \simeq R_*\bigg( 1 + \frac{t}{\tau_1}\bigg)^{2/3} \, ,
\end{equation}
where $\tau_1 \simeq R_*/v_1 = \tau_0/\zeta_*^{1/2}$ is the characteristic eddy turn-over time of the largest eddy at the beginning of the inverse cascade stage, $v_1 \simeq \zeta_*^{1/2} U_f$ is the associated characteristic plasma velocity, and we have used the relation $R_* = \tau_0 U_f$. Furthermore, the evolution of the magnetic and kinetic energy densities are given by
\begin{equation} \label{Eq:energy-density-evolution}
\begin{split}
    \rho_M(t) & \simeq \textrm{w} b_1^2\bigg( 1 + \frac{t}{\tau_1}\bigg)^{-2/3} \, ,
    \\
    \rho_K(t) & \simeq \textrm{w} v_1^2\bigg( 1 + \frac{t}{\tau_1}\bigg)^{-2/3} \, ,
\end{split}
\end{equation}
where $v_1 \simeq \zeta^{1/2}_* U_f$ and $b_1 \simeq \zeta^{1/2}_* b_0$ are, respectively, characteristic velocity and magnetic field perturbations at the beginning of the inverse cascade stage, and $v_1 \simeq b_1$ due to equipartition of the two components.
The turnover time at the correlation scale of the turbulence ($\tau_{_\xi}$) and the inverse cascade timescale ($\tau_{\rm inverse}$) then evolve as
\begin{equation} \label{Eq:cascade-time}
    \tau_{_\xi} \simeq \tau_{\rm inverse} \simeq \frac{\xi_M(t)}{v_k(t)} = \tau_1 \bigg( 1 + \frac{t}{\tau_1} \bigg) \, ,
\end{equation}   
where we have used that $v_k(t)  \propto v_1( 1 + \frac{t}{\tau_1})^{-1/3}$ from Eq.~(\ref{Eq:energy-density-evolution}). Putting Eq.~(\ref{Eq:correlation-scale}) together with Eq.~(\ref{Eq:cascade-time}), we obtain the time when turbulence exists on the scale $\xi_M(t)$ as
\begin{equation}    
    \tau_{\rm inverse} \simeq \tau_1\bigg( \frac{\xi_M(t)}{R_*} \bigg)^{3/2} = \tau_1\bigg( \frac{k_0}{k_{\xi}(t)} \bigg)^{3/2} \, ,
\end{equation}
where the wavenumber of the largest eddy is defined as $k_{\xi}(t) \equiv 2\pi/\xi_{M}(t)$.

Based on the approach used in \cite{Kahniashvili:2008er}, we compute the GW output during the inverse cascade period by adopting a stationary turbulence model wherein, rather than considering freely-decaying turbulence, we consider stationary turbulence with a duration time that depends on the scale, $k$, being considered, i.e.,
\be
    \tau_{\rm inverse}(k) \approx \tau_1\bigg(\frac{k_0}{k}\bigg)^{3/2} \,.
\ee
Thus we can express the turn-over time associated with the largest scale, $k_s$, when the inverse casade stops as
\be
    \tau_{s} \approx \tau_1\bigg(\frac{k_0}{k_s}\bigg)^{3/2} \,.
\ee
In the absence of any effective mechanisms for dissipating the turbulence at the largest scales, an inverse cascade can cause the correlation length of the magnetic field to increase greatly during this period, limited only by the Hubble expansion of the universe. Thus, the inverse cascade stops either at a scale, $\lambda_s$, when the correlation length of the turbulence reaches the Hubble radius,
\be \label{tau_s_bound0}
    \lambda_s \leq H_*^{-1} \,,
\ee
or the inverse cascade stops after a time, $\tau_s$, when the turn-over time of the largest eddy reaches the expansion timescale,
\be \label{tau_s_bound}
    \tau_s = \tau_1\bigg(\frac{k_0}{k_s}\bigg)^{3/2} 
    = \frac{\lambda_s^{3/2}}{U_f\zeta_*^{1/2}R_*^{1/2}} \leq H_*^{-1} \,.
\ee
Since $R_*H_*$, $U_f$ and $\zeta_*$ are all less than unity, we see that the inequality Eq.~\eqref{tau_s_bound} gives a stronger condition than~\eqref{tau_s_bound0}. Thus we obtain an expression for the scale at which the inverse cascade stops by saturating the inequality
\be
    \frac{\lambda_s}{R_*} \leq \bigg(\frac{U_f}{R_*H_*}\bigg)^{2/3}\zeta_*^{1/3} \,.
\ee

\subsection{Sourcing GW from turbulence}
\label{sec:sourcing}

We consider  statistically homogeneous and isotropic turbulence{, which} source{s}  GW lasting for a limited time $\tau_T < H_*^{-1}$, so that the expansion of the universe may be ignored during the period in which the gravitational radiation is produced. Furthermore, following~\cite{Gogoberidze:2007an, Kahniashvili:2008pe}, we make the additional simplifying assumption that direct cascade MHD turbulence decaying on a time scale $\tau_{\rm direct}$ is equivalent to stationary turbulence with duration $\tau_{\rm direct}/2$, as justified by the argument for unmagnetised turbulence in~\cite{Proudman:1952}. For the inverse cascade period we also consider stationary turbulence~\cite{Kahniashvili:2008pe}, but this time with a scale-dependent duration time as outlined in Section~\ref{sec:inverse}. There has been some debate in the literature \cite{Caprini:2009yp, Gogoberidze:2007an} regarding the extent to which assuming a stationary source is a valid simplification.~\footnote{This debate is especially relevant in the case of direct cascade turbulence where no attempt has been made to account for the decay of the source by considering scale dependent turbulence, in contrast to inverse cascade tubulence.} However, whilst such an approximation has limitations, it is currently the only available model that has a complete treatment of helicity and inverse cascade turbulence, and hence the potential polarisation signal. Thus, our results may be considered as a demonstration of principle, which may be used as a prototype for other, more sophisticated calculations. We expect that our results will be refined as many unknowns in the simulation and modelling of turbulence are clarified.

As shown in \cite{Gogoberidze:2007an}, in order to find the total GW energy density at a point in space and time we integrate over a spherical shell centered at that point that contains all GW sources with a light-like distance from such an observer. The thickness of the shell would then correspond to the duration of the phase transition, and its radius would correspond to the proper distance between observer and source along a light-like trajectory. {Following~\cite{Gogoberidze:2007an,Kahniashvili:2008pe} we can estimate the ensuing GW signal strength with $\pm 25 \%$ accuracy by working in the aero-acoustic approximation ($\textbf{k}\to 0$). Using our premise that the source is homogeneous and isotropic and making the aforementioned simplifying assumption that the source is stationary, the integral for the total GW energy density finally simplifies to
\be \label{rho_GW_star}
    \rho_{\rm GW}(\omega_*) = \frac{d\rho_{\rm GW}}{d\ln \omega_*} = 16\pi^3\omega_*^3G\textrm{w}^2\tau_T H_{ijij}(0,\omega_*) \, ,
\ee
where $\omega_* = \omega(t_*)$ is the angular frequency measured at the time of the phase transition and $\textrm{w}$ is {the} enthalpy density.  The scalar quantity $H_{ijij}(0,\omega_*)$ is the double trace of the four-dimensional power spectrum of the energy density tensor describing stationary turbulence in the $\textbf{k}\to 0$ approximation~\cite{Kahniashvili:2008pe}.}

\subsubsection{$H_{ijij}(0,\omega)$ behaviour}

The quantity $H_{ijij}$ controls both the peak frequency and shape of the resulting GW signal, and its functional form varies depending on whether one is considering direct cascade or inverse cascade turbulence, as we outline in more detail below.

\subsubsection*{Stage 1 - Model for the direct cascade}

In the case of direct cascade turbulence, $H_{ijij}$ takes the form~\cite{Kahniashvili:2008pe}
\be
\label{H_stage1}
    H^{(\textrm{stage} \; 1)}_{ijij}(0,\omega) 
    \approx \frac{7C_k^2\epsilon}{6\pi^{3/2}}\int_{k_0}^{k_d}\frac{dk}{k^6}\exp\bigg(-\frac{\omega^2}{\epsilon^{2/3}k^{4/3}}\bigg)\erfc\bigg({-\frac{\omega}{\epsilon^{1/3}k^{2/3}}}\bigg) 
    \,,
\ee
where $\epsilon = k_0 U_f^3 = k_0M^3$ is the energy dissipation rate per unit enthalpy, $M=U_f < 1$ is the turbulent Mach number and $C_k$ is a constant that is ${\cal O}(1)$. We assume that in the above integral $k_0 \ll k_d$, where $k_0$ is the wavenumber associated with the average bubble radius that sets the characteristic scale of the turbulence, and $k_d$ is the wavenumber associated with the scale at which the turbulence is dissipated by viscosity. As we are considering MHD turbulence, we take the prefactor of Eq.~\eqref{H_stage1} as $7/6$, after doubling the result for pure hydrodynamic turbulence given in \cite{Gogoberidze:2007an} to account for approximate equipartition between the magnetic and kinetic energy components. The integral (\ref{H_stage1}) for direct cascade turbulence is dominated by large-scale contributions at wavenumbers close to $k_0$, corresponding to the average bubble radius and, as such, we expect the GW signal for this stage of turbulence to peak at frequencies close to this scale.

\subsubsection*{Stage 2 - Model for the inverse cascade}

For the period of freely-decaying inverse cascade MHD turbulence we adopt the `Model B' introduced in Section~\ref{sec:inverse} and outlined in~\cite{Kahniashvili:2008pe}.\footnote{If we had used the `Model A' also outlined in~\cite{Kahniashvili:2008pe}, which was originally based on the work of \cite{Biskamp99,Christensson01}, we would have found the same peak frequency
but a mild suppression of the GW peak amplitude.}.
The form of $H_{ijij}(0,\omega)$ associated with the inverse cascade period is then expressed as
\bea \label{H_stage2}
   H^{(\textrm{stage} \; 2)}_{ijij}(0,\omega) &\approx  \frac{7C_1^2M^3\zeta_*^{3/2}}{6\pi^{3/2}k_0^{3/2}}\int_{k_s}^{k_0}\frac{dk}{k^{7/2}}\exp\bigg(-\frac{\omega^2k_0}{\zeta_*M^2k^3}\bigg)\erfc\bigg(-\frac{\omega k_0^{1/2}}{\zeta_*^{1/2}Mk^{3/2}}\bigg) \, ,
\eea
where $\zeta_*$ is the fraction of magnetic helicity left over at the end of the phase transition, and $C_1$ is a ${\cal O}(1)$ constant that links the magnetic energy and helicity densities with their respective power spectra. 

\section{Spectrum of the SGWB} 
\label{Spectrum_of_SGWB}

In order to calculate the spectrum of GW radiation measured today, we redshift Eq.~\eqref{rho_GW_star} to now and normalise it to the critical energy density required to make the universe flat $(k=0)$, namely $\rho_c = 3H_0^2/(8\pi G)$, defining the fraction of energy density in GWs today as
\be \label{Omega_GW,0}
     \Omega_{{\rm GW},0} = \left(\frac{a_*}{a_0}\right)^4 \frac{\rho_{{\rm GW},*}}{\rho_{c,0}} = \bigg(\frac{a_*}{a_0}\bigg)\frac{128\pi^4}{3H_0^2}\omega^3G^2\textrm{w}^2\mathlarger{\mathlarger{\sum}}_{m = 1, 2} \tau^{(\textrm{stage} \; m)}_T H^{(\textrm{stage} \; m)}_{ijij}(0,\omega_*) \,,
\ee
where $\omega = (a_*/a_0)\omega_*$, the enthalpy density $\textrm{w} = 4\rho_*/3 = 2\pi^2 g_*T_*^4/45$, and $H_*^2 = 8\pi G\rho_*/3 = 8\pi^3 G g_*T_*^4/90$. Rearranging this relation, we obtain $G^2\textrm{w}^2 = H_*^4/4\pi^2$, and substituting this back into the Eq.~\eqref{Omega_GW,0} {we get
\bea \label{Omega_GW}
   \Omega_{{\rm GW},0} &= \bigg(\frac{a_*}{a_0}\bigg)\frac{32\pi^2}{3H_0^2}\omega^3H_*^4\mathlarger{\mathlarger{\sum}}_{m = 1, 2} \tau^{(\textrm{stage} \; m)}_T H^{(\textrm{stage} \; m)}_{ijij}(0,\omega_*) 
\\
    &= \bigg(\frac{a_*}{a_0}\bigg)\frac{1 \times 10^{37}}{\textrm{Hz}^2}\omega^3H_*^4 \mathlarger{\mathlarger{\sum}}_{m = 1, 2} \tau^{(\textrm{stage} \; m)}_T H^{(\textrm{stage} \; m)}_{ijij}(0,\omega_*) \, ,
\eea
where we have used $H_0 = h_0 \times 100 \:\textrm{km}\, \textrm{sec}^{-1} \, \textrm{Mpc}^{-1}$ with $h_0 = 0.67$~\cite{Aghanim:2018eyx}.}


\subsection{Stage 1 - direct cascade} \label{sec:stage1}

\subsubsection{Stationary approximation}

After normalising Eq.~\eqref{H_stage1}, we can express the peak frequency of the GW signal due to direct cascade turbulence at the time of the transition as
\be \label{tina_peak}
    f_{{\rm peak},*} = 1.48M/R_* \,,
\ee
which after red-shifting gives a peak frequency today of
\be \label{peak_frequency_stage1}
    f_{{\rm peak},0} = \frac{a_*}{a_0} f_{{\rm peak},*}  = 2.45 \times 10^{-5} \,{\rm Hz}\, \left(\frac{T_*}{100\,{\rm GeV}} \right) \bigg(\frac{g_*}{100}\bigg)^{1/6}\frac{M}{R_*H_*} \, ,
\ee
where we have used the relation 
\be \label{a_star_over_a_0}
    \frac{a_*}{a_0} \approx 8 \times 10^{-16}\bigg( \frac{100 \textrm{GeV}}{T_*} \bigg)\bigg(\frac{100}{g_*}\bigg)^{1/3} \, .
\ee
The peak amplitude of the direct cascade GW signal can then be written as
\be \label{peak_amplitude_stage1}
   \Omega_{{\rm GW},0}^{(\rm stage~1)} = 7.357 \times 10^{-6} \bigg(\frac{100}{g_*}\bigg)^{1/3}(R_*H_*)\big(\tau_T H_*\big)C_k^2M^6 \,,
\ee
where $\tau_T H_* \leq 1$ is the duration of the Stage 1 direct cascade turbulence normalised to the Hubble time, $R_* H_*$ is the average bubble radius at percolation normalised to the Hubble radius, $M$ is the Mach number and $C_k$ is a constant of order unity. 

The $\tau_T H_*$ factor in Eq.~\eqref{peak_amplitude_stage1} tells us that, as expected, the longer lasting the period of direct cascade turbulence the larger the abundance of GW emitted during this direct cascade stage. The average bubble size $R_*H_* < 1$ sets the characteristic length scale of the problem, and thereby controls the peak frequency of the GW spectrum arising from direct cascade turbulence, as seen in Eq.~\eqref{peak_frequency_stage1}. A larger bubble radius $R_*H_*$ also implies fewer bubbles per Hubble horizon, which in turn means a higher energy concentration as the bubbles convert vacuum energy from their volume into the walls. After the bubble collisions this results in a more inhomogeneous energy distribution centred around the scale $R_*$, and thus a higher abundance of GWs as exhibited by the factor $\propto R_*H_*$ seen in~\eqref{peak_amplitude_stage1}.

\subsubsection{Other models for direct cascade turbulence}

Several other models for approximating the GW signal from direct cascade turbulence have been proposed in
the literature. Generalising~\eqref{tina_peak}, they may be characterised by the peak frequency at the time of the phase transition: 
\be
    f_{\textrm{peak},*} = \frac{A}{R_*} \, ,
\ee
which is redshifted to the following generalisation of~\eqref{peak_frequency_stage1} today,
\be \label{eq:f_peak}
    f_{\textrm{peak},0} = B\, \textrm{Hz} \,\frac{T_*}{100 \textrm{GeV}}\bigg(\frac{g_*}{100}\bigg)^{\frac{1}{6}}\frac{1}{R_*H_*} \,,
\ee
where $A$ and $B$ are constants that depend on the way the turbulent GW source is modelled. They take the following
values in some commonly-used source models: 
\begin{itemize}
    \item The stationary approximation discussed above yields $A=1.48 M$, $B=2.45 \times 10^{-5} M$, and the spectra shown as {black} curves in {Figs.~\ref{fig:GW_tot_spectra_fixed_temp} \& \ref{fig:GW_tot_spectra_vary_temp}} below;
    \item The top-hat approximation yields $A=5.1$, $B= 8.46 \times 10^{-5}$ (see Eq.~(85) of \cite{Caprini:2009yp}), and the spectra shown as {dark grey} curves in {Figs.~\ref{fig:GW_tot_spectra_fixed_temp} \& \ref{fig:GW_tot_spectra_vary_temp}} below;
    \item The coherent approximation yields $A=0.586$, $B = 9.728 \times 10^{-6}$ (see Eq.~(80) of~\cite{Caprini:2009yp});
    \item The incoherent approximation yields $A=8.64$, $B = 1.43 \times 10^{-4}$ (see Eq.~(76) of \cite{Caprini:2009yp}).
\end{itemize}

In the LISA phase transitions working group review paper~\cite{Caprini:2015zlo} the main source of GW signal from plasma flow is associated with sound waves. The GW spectrum for this source is~\cite{Hindmarsh:2019phv,Hindmarsh:2017gnf,Hindmarsh:2016lnk,Hindmarsh:2015qta}
\be
   h^2\Omega_{{\rm sw},0} = 0.9 \times 10^{-6}\left(R_*H_*\right)\left(\tau_{\rm sw} H_*\right)\bigg(\frac{\kappa_{\textrm{v}}\alpha}{1+\alpha}\bigg)^2\bigg( \frac{100}{g_*} \bigg)^{\frac{1}{3}}S_{\textrm{sw}}(f) \,,
\ee
where $\kappa_{\rm sw}$ is the efficiency with which vacuum energy is transformed into bulk motion of the fluid (and can be easily expressed for fast bubble walls, see Eq.~\eqref{eq:kappa_sw}), and the spectral shape is
\be
    S_{\textrm{sw}}(f) = (f/f_{sw})^3\bigg( \frac{7}{4 + 3(f/f_{sw})^2 } \bigg)^{7/2} \, .
\ee
Finally, there is an additional suppression factor that depends on fluid velocity (see Eq.~\eqref{eq:U_f}) 
\be
    \tau_{\rm sw} H_* = \min \left( 1, \frac{R_* H_*}{U_f}  \right) \, ,
\ee
which is associated with the time at which shocks develop in the flow~\cite{Hindmarsh:2017gnf}, and is much less than one for most models~\cite{Ellis:2020awk,Ellis:2019oqb,Caprini:2019egz,Ellis:2018mja}. The peak frequency of the sound wave source at the time of the phase transition reads
\be
    f_{\textrm{sw},*} = \frac{3.38}{R_*} \, 
\ee
which becomes 
\be
    f_{\textrm{sw},0} = 5.61 \times 10^{-5}\,{\rm Hz}\, \frac{T_*}{100 \textrm{GeV}}\bigg(\frac{g_*}{100}\bigg)^{\frac{1}{6}}\frac{1}{R_*H_*} 
\ee
when redshifted to the present day. This contribution provides the {light grey} curves in {Figs.~\ref{fig:GW_tot_spectra_fixed_temp} \& \ref{fig:GW_tot_spectra_vary_temp}} below.

{In order to model the GWs sourced from turbulence, Ref.~\cite{Caprini:2015zlo} used the top-hat approximation to estimate the signal for reasons outlined in~\cite{Caprini:2009yp}.}
Assuming Kolmogorov-type turbulence, they calculate the associated GW spectrum arising from this model to be
\be
   h^2\Omega_{{\rm turb},0} = 1.14 \times 10^{-4} \left(R_*H_*\right) \bigg(\frac{\kappa_{\textrm{sw}}\alpha}{1+\alpha}\bigg)^{\frac{3}{2}}\bigg( \frac{100}{g_*} \bigg)^{\frac{1}{3}}S_{\textrm{turb}}(f) \,,
\ee
where $v_w$ is the wall velocity~\footnote{We assume $v_w \sim 1$ in this work.}, $R_*H_*$ is the amplitude suppression factor discussed in the previous section, and we have also used $\kappa_{\rm sw}$ as the efficiency for conversion of the latent heat released during the phase transition into MHD turbulence. This comes from our optimistic assumption that when the flow becomes non-linear and the sound wave period ends, the remaining energy is readily converted into turbulence. Given that we discuss scenarios in which the sound wave period lasts a relatively short time, very little energy is lost and we expect this to be a reasonable approximation. However, in principle there can be an extra damping factor due to, for example, loss of sound wave energy into reheating the plasma. The corresponding spectral shape is
\be
    S_{\textrm{\textrm{turb}}}(f) =  \frac{(f/f_{\textrm{turb}})^3}{[1 + (f/f_{\textrm{turb}})]^{\frac{11}{3}}(1+8\pi f/h_*) } \,,
\ee
where 
\be
    h_* = 16.5 \times 10^{-6} \, \textrm{Hz} \, \bigg(\frac{T_*}{100 \textrm{GeV}}\bigg)\bigg(\frac{g_*}{100}\bigg)^{1/6} 
\ee
is the inverse Hubble time at GW production redshifted to today. This contribution provides the {dark grey} curves in {Figs.~\ref{fig:GW_tot_spectra_fixed_temp} \& \ref{fig:GW_tot_spectra_vary_temp}} below.

\subsection{Stage 2 - inverse cascade} 
\label{sec:stage2}

In contrast to the direct cascade, the length scale providing the largest contribution to the $H^{(\textrm{stage} \; 2)}_{ijij}(0,\omega)$ {quantity} used to calculate the GW signal arising from the inverse cascade period of  turbulence is model-independent, being simply set by the Hubble scale. This is because the majority of the turbulent energy is found around the Hubble scale at the end of the inverse cascade before it is dissipated due to the expansion of the universe as outlined in Section~\ref{sec:inverse}. Taking the value of the Hubble parameter at the phase transition and red-shifting it to today, we find that the characteristic frequency of the inverse cascade GW spectrum today is
\be \label{eq:f_horizon}
    f_{{\rm horizon},0} = 1.65 \times 10^{-5}\, \textrm{Hz}\, \bigg( \frac{T_*}{100 \textrm{GeV}} \bigg) \bigg(\frac{g_*}{100}\bigg)^{1/6} \,.
\ee
As we will see in the examples in the next section, at this frequency the power-law of abundance of GWs changes from $\Omega_{\rm GW}\propto f^3$ as expected beyond the horizon scale~\cite{Cai:2019cdl} to a flatter plateau composed of contributions from both the direct cascade turbulence and the helicity fraction dependent inverse cascade turbulence. The size of the plateau depends on the magnitude of the helicity fraction: for small $\zeta_*$ the signal briefly levels off before reverting to its original $\Omega_{\rm GW}\propto f^3$ growth rate; whilst for sufficiently large $\zeta_*$ the plateau continues all the way up to the scale associated with the bubble size at the transition (see Eq.~\eqref{eq:f_peak}).

\subsection{GW spectrum plots}

\begin{figure}
    \centering
    \includegraphics[width=0.99\textwidth]{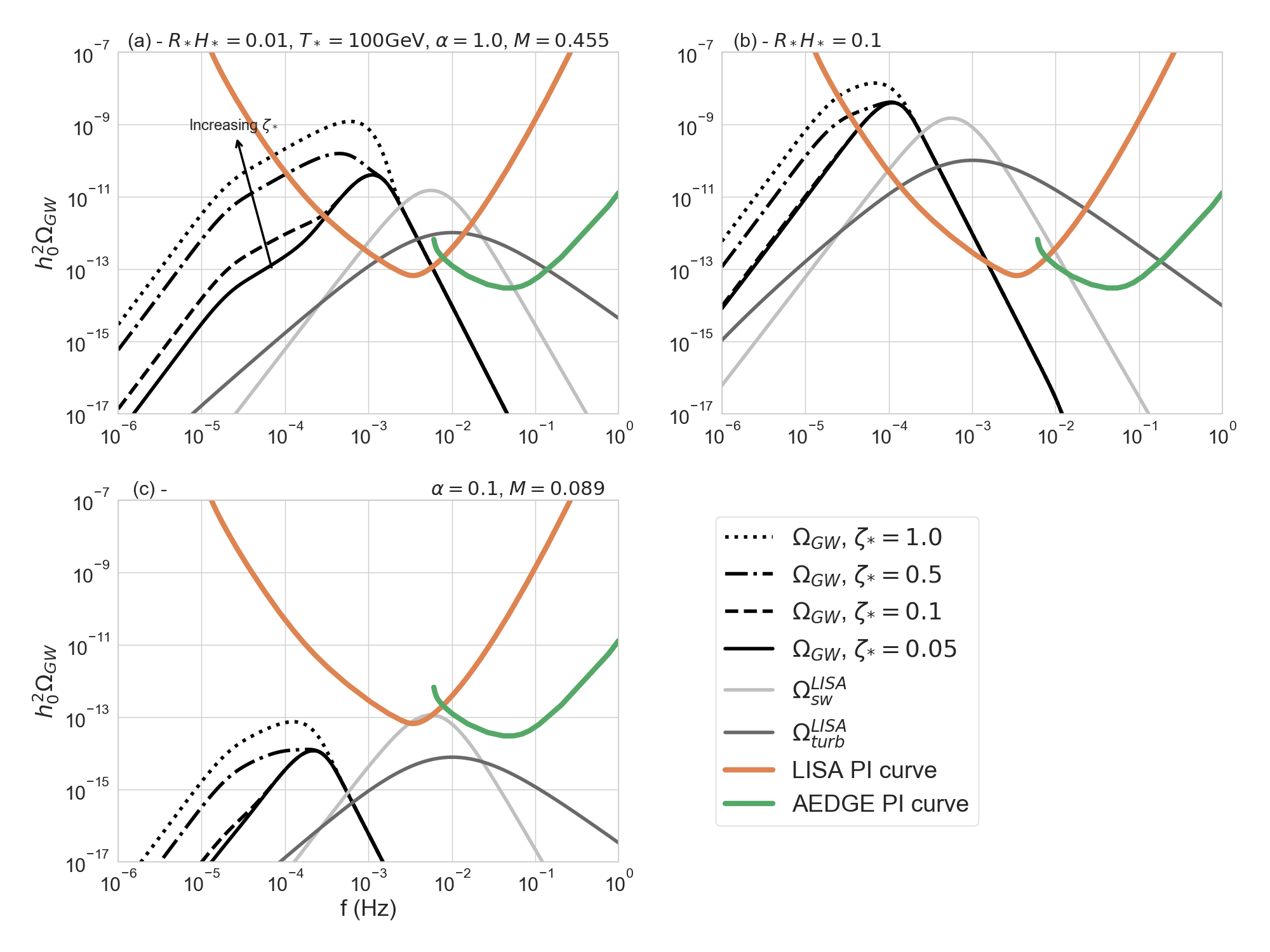}
    \caption{\it {Our calculations of GW spectra for fixed $T_* = 100$~GeV 
    and $\zeta_* = 0.05$ (solid {black}), $0.1$ (dashed {black}), $0.5$ (dash-dotted {black}) and $1.0$ (dotted {black}), compared with calculations of the spectra from sound waves ({light grey}) and turbulence ({dark grey}) taken from~\cite{Caprini:2015zlo}. The value of $R_*H_*$ increases from left to right and the value of $\alpha$ decreases from top to bottom. The parameter values are the same as in panel (a) unless specified. The {power-law integrated (PI)} LISA~\cite{Caprini:2015zlo} sensitivity to the total SGWB spectrum is shown as an {orange} line and the PI AEDGE~\cite{AEDGE} sensitivity is shown as a {green} line, {each for a 4-year integration time}.}}
    \label{fig:GW_tot_spectra_fixed_temp}
\end{figure}

\begin{figure}
    \centering
    \includegraphics[width=0.99\textwidth]{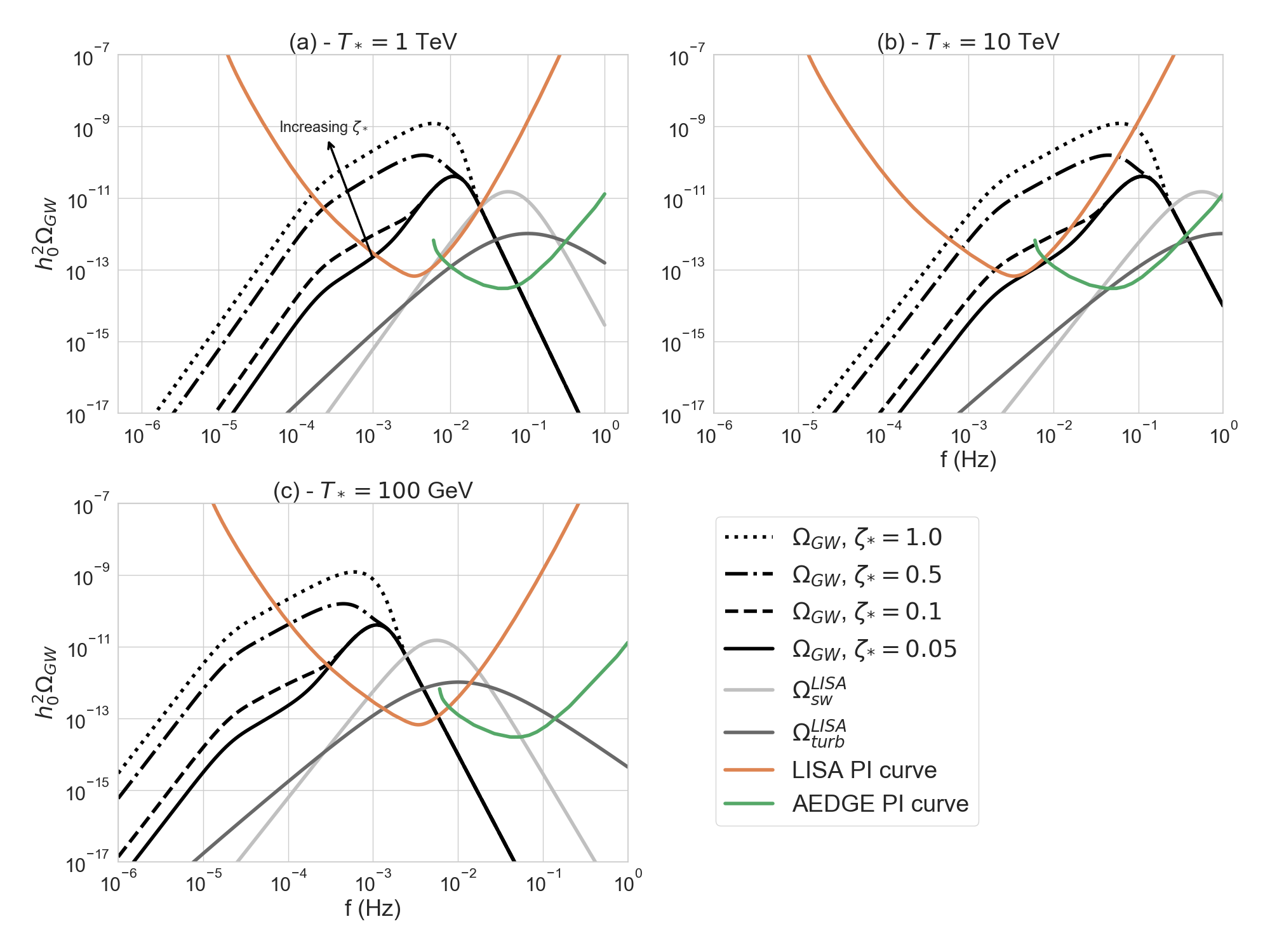}
    \caption{\it{{Similar to Fig~\ref{fig:GW_tot_spectra_fixed_temp} but for fixed $\alpha=1.0$, $R_*H_* = 0.01$ and varying $T_*$, which increases from left to right and decreases from top to bottom.}}}
    \label{fig:GW_tot_spectra_vary_temp}
\end{figure}

{Fig.~\ref{fig:GW_tot_spectra_fixed_temp} and Fig.~\ref{fig:GW_tot_spectra_vary_temp} compare our calculated GW spectra (black) for representative choices of the model parameters $\alpha, R_*$ and $T_*$ with the LISA~\cite{Caprini:2015zlo} and AEDGE~\cite{AEDGE} sensitivity curves (shown in orange and green, respectively). The value of $M$ is not an independent quantity, being related to the magnitude of $\alpha$ (see Eq.~\eqref{eq:U_f}). Our calculations are for four values of the helicity fraction $\zeta_* = 0.05$ (solid), $0.1$ (dashed), $0.5$ (dash-dotted) and $1$ (dotted).}

{The comparisons in Fig.~\ref{fig:GW_tot_spectra_fixed_temp} are for fixed $T_* = 100$~GeV and different choices of {the parameters} $\alpha$ (going down) and $R_*H_*$ (going across), describing the strength of the first-order transition and the bubble size $R_*$ respectively. We recall that $\alpha$ sets the value of the turbulent Mach number M through Eq.~\eqref{eq:U_f}. The value of M affects the peak frequency and amplitude of the direct cascade GW signal through Eqs.~(\ref{peak_frequency_stage1}) and (\ref{peak_amplitude_stage1}), respectively, whilst the characteristic amplitude of the inverse cascade GW signal is given by Eq.~\eqref{H_stage2}.}

{Fig.~\hyperref[fig:GW_tot_spectra_fixed_temp]{1(a)} shows that, as $\zeta_*$ increases, the size of the low-frequency plateau in the GW spectrum arising from the inverse cascade period of turbulence, which was discussed in Section~\ref{sec:stage2}, also increases. Indeed, for large enough $\zeta_*$ where the contribution to the GW spectrum from the inverse cascade period is sufficiently sizeable, the plateau transitions into a distinct new peak at a frequency slightly below the frequency of the direct cascade peak. Similar features are seen in Fig.~\hyperref[fig:GW_tot_spectra_fixed_temp]{1(b)} and \hyperref[fig:GW_tot_spectra_fixed_temp]{1(c)}.}

{Comparing Fig.~\hyperref[fig:GW_tot_spectra_fixed_temp]{1(a)} with Fig.~\hyperref[fig:GW_tot_spectra_fixed_temp]{1(b)}, we see that decreasing the value of $R_*H_* \leq 1$ both suppresses the amplitude of the GW signal and pushes it to higher frequencies. This is to be expected from the analysis in Section~\ref{sec:stage1}. Furthermore, we see that the relative contribution of the low-frequency inverse cascade turbulence to the overall GW amplitude decreases with increasing $R_*H_*$, because for larger values of $R_*$ the inverse cascade turbulence has less time to develop before being washed out by the Hubble expansion.}

{Comparing Fig.~\hyperref[fig:GW_tot_spectra_fixed_temp]{1(a)} with Fig.~\hyperref[fig:GW_tot_spectra_fixed_temp]{1(c)} where the value of $\alpha$ (and thus the value of M) has been decreased, we see that for smaller values of $\alpha$ the peak frequency of the GW spectrum is shifted to lower frequencies and the amplitude is suppressed.}

{Comparing our predictions with the top-hat approximation favoured in the LISA phase transitions working group review paper~\cite{Caprini:2015zlo} (dark grey curves), we see that the peak frequencies are closer for larger $\alpha$ (and $M$). The heights of our peaks increase with $\zeta_*$ and are generally higher than the top-hat peaks for $\alpha = 1.0$, but lower for $\alpha = 0.1$. Both our calculations and the top-hat approximation for the choices $\alpha = 1.0$ and $R_* H_* = 0.01$ (Fig.~\hyperref[fig:GW_tot_spectra_fixed_temp]{1(a)}) and $0.1$ (Fig.~\hyperref[fig:GW_tot_spectra_fixed_temp]{1(b)}) yield spectra peaking well within the sensitivity of LISA, whereas for $\alpha = 0.1$ and $R_* H_* = 0.01$ (Fig.~\hyperref[fig:GW_tot_spectra_fixed_temp]{1(c)}) both peaks lie below the LISA sensitivity.}

{In Fig.~\ref{fig:GW_tot_spectra_vary_temp} we display comparisons similar to Fig.~\ref{fig:GW_tot_spectra_fixed_temp}, but now fixing $\alpha = 1.0$ and $R_* H_* = 0.01$ and choosing different transition temperatures $T_*$.
(Fig.~\hyperref[fig:GW_tot_spectra_fixed_temp]{1(a)} is repeated here as panel (c).)
The peaks of the calculated spectra shift to larger frequencies for larger $T_*$. We can see from Fig.~\hyperref[fig:GW_tot_spectra_vary_temp]{1(a)} that our calculations for $T_* = 1$~TeV peak within the LISA~\cite{Caprini:2015zlo} sensitivity, whereas the peak of the top-hat calculation peaks within the AEDGE~\cite{AEDGE} sensitivity. Fig.~\hyperref[fig:GW_tot_spectra_vary_temp]{1(b)} shows that for $T_* = 10$~TeV our peak reaches within the AEDGE sensitivity, whereas the peak of the top-hat calculation peaks at higher frequency.}

{Our calculations indicate that LISA and AEDGE have complementary capabilities to detect the SGWB from a first-order phase transition, with the higher frequency range of AEDGE extending the detectable range of $T_*$ to higher values.}

\section{Circular polarisation of the SGWB}
\label{sec:polarisation}

\subsection{Computation of the polarised GW spectra}
\label{sec:computation}

The circular polarisation of a GW signal is given by~\cite{Kahniashvili:2005qi,Kisslinger:2015hua}
\be \label{pol_frac}
\mathcal{P}_{\rm GW}(k)=\frac{\left\langle h^{+\star}(\mathbf{k}) h^{+}\left(\mathbf{k}^{\prime}\right)-h^{-\star}(\mathbf{k}) h^{-}\left(\mathbf{k}^{\prime}\right)\right\rangle}{\left\langle h^{+\star}(\mathbf{k}) h^{+}\left(\mathbf{k}^{\prime}\right)+h^{-\star}(\mathbf{k}) h^{-}\left(\mathbf{k}^{\prime}\right)\right\rangle} = \frac{\mathcal{I}_A(K)}{\mathcal{I}_S(K)} 
\,,
\ee
where $h^+$ and $h^-$ are the states corresponding to right- and left-handed circularly polarised GWs, and $K=k/k_0$ is a wavenumber normalised to the wavenumber associated with the bubble radius at collision, $k_0$. The explicit calculation of the polarisation of the GW spectrum using Eq.~\eqref{pol-integrals} is only required for the direct cascade period of turbulence where $\zeta_* < 1$. For the GW spectrum emitted during the inverse cascade period, on the other hand, we simply assume that the emitted GW spectrum is fully polarised, on the premise that $\zeta_* \simeq 1$ at the beginning of the inverse cascade stage.

In the particular case of helical turbulence the relevant functions can be approximated as~\cite{Kisslinger:2015hua}
\bea \label{pol-integrals}
\mathcal{I}_S(K) & \simeq \int d P_{1} \, P_{1} \int d P_{2} \, P_{2} \,\bar{\Theta}\left[\left(1+\gamma_{p}^{2}\right)\left(1+\beta_{p}^{2}\right) P^{n_S}_{1}P^{n_S}_{2} +4 h^2 \gamma_{p} \beta_{p} P^{n_A}_{1}P^{n_A}_{2}\right] \,, \\
\mathcal{I}_A(K)  & \simeq 2 h \int d P_{1} \, P_{1} \int d P_{2} \, P_{2} \,\bar{\Theta}\left[\left(1+\gamma_{p}^{2}\right) \beta_{p} P^{n_S}_{1}P^{n_A}_{2} + \left(1+\beta_{p}^{2}\right) \gamma_{p}P^{n_A}_{1} P^{n_s}_{2}\right] \, ,
\eea
where
\be
\begin{aligned}
\gamma_{p} & =\frac{K^{2}+P_{1}^{2}-P_{2}^{2}}{2 K P_{1}}, \quad \quad \beta_{p}= \frac{K^{2}+P_{2}^{2}-P_{1}^{2}}{2 K P_{2}}, \\
\bar{\Theta} & = \theta\left(P_{1}+P_{2}-K\right) \theta\left(P_{1}+K-P_{2}\right) \theta\left(P_{2}+K-P_{1}\right) \, , 
\end{aligned}
\ee
and $\theta$ is the Heaviside step function. The parameter $h$ is the fraction of helicity dissipation as defined in {\cite{Kisslinger:2015hua}}, which is related to the magnetic helicity fraction. Indeed, these two parameters coincide in the helical Kolmogorov turbulence model: $\zeta_* \simeq h$~{\cite{Kisslinger:2015hua}}. Following the approach in \cite{Kisslinger:2015hua}, which seeks to generalise the polarisation degree calculation for helical hydrodynamic turbulence given in~\cite{Kahniashvili:2005qi} to the case of helical MHD turbulence, in this paper we determine the polarisation degree of the direct cascade period by modelling the source as stationary\footnote{Debate regarding the validity of modelling the direct cascade source as stationary is of less importance when the helicity fraction is large given our results shown later demonstrate that in this regime the inverse cascade spectrum, which attempts to account for the turbulent decay through scale dependent decorrelation, dominates the overall signal.} and using the symmetric and helical spectral indices, $n_S=-11/3$ and $n_A = -14/3$, consistent with a helical Kolmogorov spectrum.\footnote{If the helicity fraction is large, strong helical turbulence modelling~\cite{Kahniashvili:2005qi} could be more appropriate, and would result in a larger helicity fraction from the first stage of turbulence. However, as we demonstrate later and have explicitly checked for all our results, the contribution from the second stage of turbulence is always clearly dominant in the case of a large helicity fraction, and for simplicity we use here the Kolmogorov spectrum also for the first stage.} We take the integration limits in Eq.~\eqref{pol-integrals} to range from $1$ to $k_d/k_0$, and simply discard scales larger than the bubble radius, i.e., $k<k_0$, which are only relevant to the inverse cascade period of turbulence.

\begin{figure}
    \centering
    \includegraphics[width=0.6\textwidth]{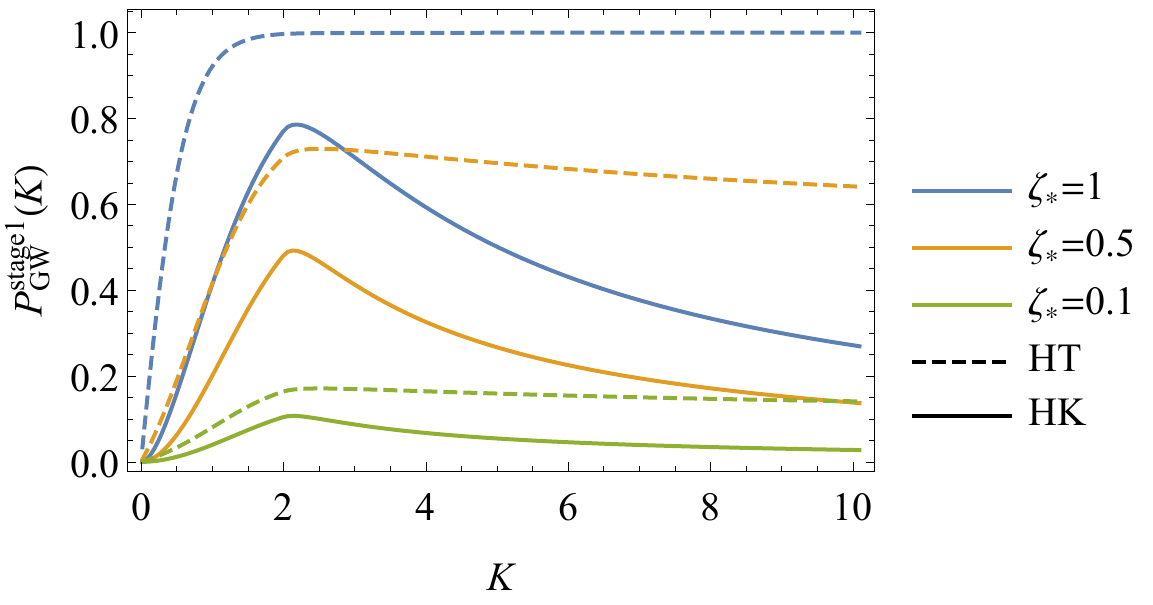}
    \caption{\it The degree of polarisation of Stage 1 direct cascade GWs as a function of the normalised wavenumber $K=k/k_0$, assuming the indicated values of the helicity dissipation parameter $h$. The value of $h$ coincides with the initial magnetic helicity fraction $\zeta_*$ for the helical Kolmogorov turbulence (HK) model as considered in this paper, shown by solid lines. We also show with dashed lines the polarisation fraction for Stage 1 turbulence driven by the Helicity Transfer (HT) model whose use may be more appropriate in the large helicity regime. However, we find the choice of Stage 1 model has no material impact on our overall results in this scenario as discussed in the text.}
    \label{fig:pol_degree}
\end{figure}

Using $\Omega_{\rm GW}(k) \propto k^5 \langle h(k)^2 \rangle$, we can rearrange Eq.~\eqref{pol_frac} to obtain
\be \label{eq:PGW}
    \Omega_{\textrm{\rm GW}}(k)\mathcal{P}_{\rm GW}(k) = \ \Omega^+_{\rm GW}(k) - \Omega^-_{\rm GW}(k) \, ,
\ee
where $\Omega_{\rm GW}(k) = \Omega^+_{\rm GW}(k) + \Omega^-_{\rm GW}(k)$. Then, for the helical turbulence model we consider in this paper, we have
\bea \label{pol_spectra}
    \Omega^+_{\rm GW}(k) & = \frac{1 + \mathcal{P}_{\rm GW}^{\textrm{stage 1}}(k)}{2}\Omega_{\textrm{stage 1}}(k) + \Omega_{\textrm{stage 2}}(k) \,,\\
    \Omega^-_{\rm GW}(k) & = \frac{1 - \mathcal{P}_{\rm GW}^{\textrm{stage 1}}(k)}{2}\Omega_{\rm stage \; 1}(k) \, ,
\eea
where we have assumed $\mathcal{P}_{\rm GW}^{\textrm{stage 2}}(k) \approx 1$, on the basis that by the beginning of the second stage $\zeta_* \approx 1$ and we are in a regime of strong helical turbulence that can be well approximated by a helicity transfer spectrum \cite{Kahniashvili:2005qi, Kisslinger:2015hua} where $\mathcal{P}_{\rm GW}(k) \approx 1$ across the range of $k$ relevant to inverse cascade turbulence. In the large wave-number limit where this approximation could no longer hold, the contribution of Stage 2 to the GW abundance is negligible. 
 {We plot the degree of polarisation of GWs emitted during Stage 1 direct cascade turbulence in Fig.~\ref{fig:pol_degree}.
We see that $\mathcal{P}_{\rm GW}^{\textrm{stage 1}}$ reaches a peak at $K \sim 2$, whose height increases
with $h = \zeta_*$, and then falls for larger $K$.}

\begin{figure}
    \centering
    \includegraphics[width=17cm]{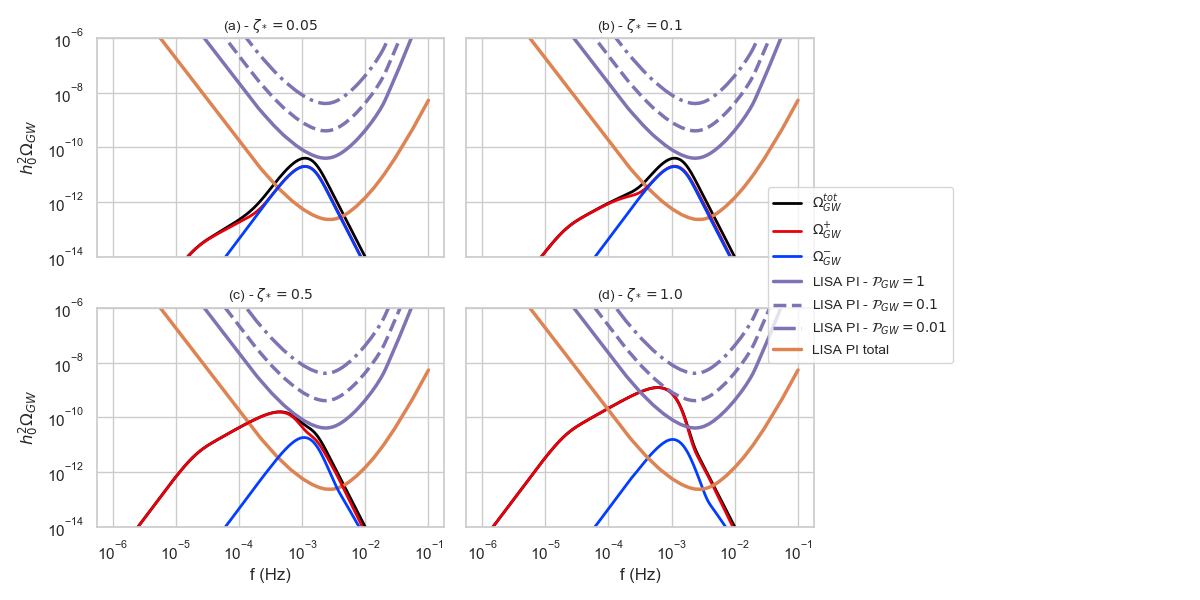}
    \caption{{ {\it The strengths of the polarised $\Omega_{GW, \pm}$ signals for various values of the initial helicity fraction, $\zeta_*$ and fixed $\alpha = 1.0$, $T_* = 100$~GeV and $R_*H_*=0.01$. The {orange} curves show the power-law integrated sensitivity of LISA for the total GW signal, and the purple lines show LISA's power-law integrated sensitivity to the polarised signal assuming polarisation fractions (see Eq.~\eqref{eq:PGW}) $\mathcal{P}_{GW}=1, \, 0.1$ and $0.01$ as the solid, dashed and dot-dashed lines, respectively.}}}
    \label{fig:pol}
\end{figure}

Fig.~\ref{fig:pol} displays the strengths of the signals for different GW polarisations $\Omega_{\rm GW}^{\pm}$ for fixed $T_* = 100$~GeV, $R_*H_*=0.01$, $\alpha = 1.0$, and various choices of the initial helicity fraction, $\zeta_*$. We {see in Fig.~\hyperref[fig:pol]{4(a)}} that for small $\zeta_* \lesssim 0.05$ the total GW signal $\Omega_{\rm GW}^{\rm tot}$ is dominated by the contribution from direct cascade turbulence with negligible net polarisation, i.e., $\Omega_{\rm GW}^{+} \simeq \Omega_{\rm GW}^{-}$. Conversely, the small low-frequency inverse cascade plateau in the signal emits fully-polarised GW, $\Omega_{\rm GW}^{+} \simeq \Omega_{\rm GW}^{\rm tot}$, as expected from our previously-stated assumption that $\zeta_* \simeq h$. Moving to {Fig.~\hyperref[fig:pol]{4(b)}}, we see that raising $\zeta_*$ to $0.1$ increases the size of the fully-polarised inverse cascade plateau in the GW signal, an effect that continues until $\zeta_* \sim 0.5$ ({Fig.~\hyperref[fig:pol]{4(c)}}) where it begins to dominate and transitions from being a plateau into a distinct new peak 
of the total GW signal. We see in the lower two panels of Fig.~\ref{fig:pol} that for $\zeta_* \gtrsim 0.5$ the contribution of the fully-polarised inverse cascade GW increasingly dominates that of the total GW signal, $\Omega_{\rm GW}^{\rm tot}$.

\begin{figure}
    \centering
    \includegraphics[width=17cm]{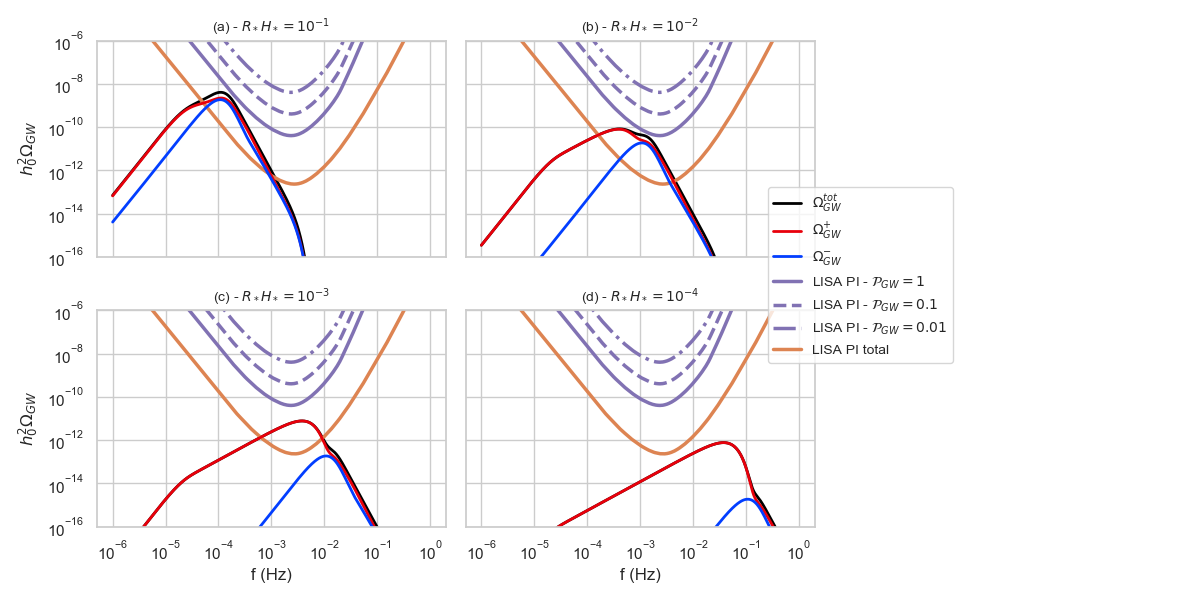}
    \caption{{\it The strengths of the polarisation of the $\Omega_{GW, \pm}$ signals for various values of the average bubble radius, $R_*H_*$, $\zeta_*$ and fixed $\zeta_* = 0.4$ $\alpha = 1.0$ and $T_* = 100$~GeV. The orange and purple sensitivity curves are the same as in Fig~\ref{fig:pol}.}}
    \label{fig:pol_vary-RH}
\end{figure}

Fig.~\ref{fig:pol_vary-RH} shows the $\Omega_{\rm GW}^{\pm}$ signal strengths for fixed $T_* = 100$~GeV, $\zeta_*=0.4$, $\alpha = 1.0$,  and various choices of $R_*H_*$. We can see that as $R_*H_*$ decreases the total GW amplitude decreases, as expected from the discussion in the previous section. However, the net polarisation of the signal increases as $R_*H_*$ decreases. This is because inverse cascade turbulence is more important at smaller $R_*H_*$ since the turnover time of the largest eddy, $\tau_s$, takes longer to reach the Hubble timescale for phase transitions with smaller average bubble radius. 
Thus the duration of the inverse cascade stage where fully-polarised GWs are emitted increases, resulting in a GW signal with larger net polarisation.

{Figs.~\ref{fig:pol} and \ref{fig:pol_vary-RH} also feature power-law integrated sensitivities for LISA: the {orange} curves are the usual PI sensitivity for the total gravitational wave signal~\cite{Thrane:2013oya}, while in purple we show PI curves for a polarisation signal. We refer the reader to Section~\ref{sec:SGWBmeasurement} for a formal derivation.
The interpretation is the same as in the unpolarised case, i.e., a power law with a polarisation fraction $\mathcal{P}_{GW}$ (see Eq.~\eqref{eq:PGW}) crossing a purple line with the same $\mathcal{P}_{GW}$ gives  SNR$\geq 10$ for a polarisation measurement with LISA.
}

As can be seen {in Fig.~\ref{fig:GW_tot_spectra_vary_temp}}, the effect of increasing $T_*$ would be to shift the $\Omega_{\rm GW}$ signal to higher frequencies, without changing the relative amounts of $\Omega_{\rm GW}^{\pm}$ or their dependences on $\zeta_*$ and $R_*H_*$.

\subsection{Measurements of the SGWB and its polarisation}
\label{sec:SGWBmeasurement}

The signal-to-noise ratio (SNR) for combining two GW detector channels $O$ and $O'$ is 
\be
{\rm SNR}_{O O'} = \sqrt{\int_0^{t_{\rm obs}} \td t \int \td f \frac{S_{O O'}^*(f) S_{O O'}(f)}{P_{n,O}(f) P_{n,O'}(f)}} \;,
\ee
where $t_{\rm obs}$ is the total duration of the measurement and $P_{n,O}(f)$ and $P_{n,O'}(f)$ are the noise spectral functions in the channels $O$ and $O'$. Following Ref.~\cite{Domcke:2019zls}, the signal function $S_{O O'}(f)$ for a stochastic GW background, defined via the Fourier transform of the correlator of the GW signals passing through the $O$ and $O'$ channels, can be expanded as a function of the peculiar velocity of the solar system $v=1.23\times 10^{-3}$ as
\be
S_{O O'}(f) = \frac{3 H_0^2}{8\pi^2 f^3} \sum_\lambda\! \left\{\! \mathcal{M}_{O O'}^\lambda(f) \, \Omega_{\rm GW}^\lambda(f) - 4i\, v \, \mathcal{D}_{O O'}^\lambda(f)  \left[\Omega_{\rm GW}^\lambda(f) \!-\! \frac{f}{4} \frac{\td \Omega_{\rm GW}^\lambda(f)}{\td f}\right] + \mathcal{O}(v^2) \!\right\} \,,
\ee
where $\lambda=\pm 1$ is the GW helicity and
\bea
&\mathcal{M}_{O O'}^\lambda(k) = 4\int \frac{\td \Omega_k}{4\pi} e_{ab,\lambda}(\hat{k})  e_{cd,\lambda}(-\hat{k}) \mathcal{Q}_O^{ab}(\vec{k}) \mathcal{Q}_{O'}^{cd}(-\vec{k}) \,, \\
&\mathcal{D}_{O O'}^\lambda(k,\hat{v}) = 4i \int \frac{\td \Omega_k}{4\pi} e_{ab,\lambda}(\hat{k})  e_{cd,\lambda}(-\hat{k})  \mathcal{Q}_O^{ab}(\vec{k}) \mathcal{Q}_{O'}^{cd}(-\vec{k}) \, \hat{k}\cdot \hat{v} \, ,
\eea
are the monopole and dipole response functions and $k \equiv |\vec{k}| = 2\pi f$. The matrices $\mathcal{Q}_{O,O'}^{ab}(\vec{k})$ contain the geometries of the detector channels, and the product of the polarisation operators appearing in the above expressions is given by
\be
e_{ab,\lambda}(\hat{k})  e_{cd,\lambda}(-\hat{k}) = \frac{1}{4}\left( \delta_{ac} - \hat{k}_a \hat{k}_c - i\lambda\epsilon_{ace} \hat{k}^e \right) \left( \delta_{bd} - \hat{k}_b \hat{k}_d - i\lambda\epsilon_{bde} \hat{k}^e \right) \,.
\ee

LISA consists of three spacecraft arranged in an equilateral triangle whose sides provide three baselines, and the combinations of pairs of these baselines form three laser interferometers. We denote the positions of the vertices of the triangle by $\vec{x}_i$ and we fix the side length to $L = 2.5\times 10^6\,$km~\cite{Audley:2017drz}. Linear combinations of the three interferometers can be used to construct three detector channels: $A$, $E$ and $T$~\cite{Adams:2010vc}. In the following we focus on the $A$ and $E$ channels, which are antisymmetric combinations of the three interferometers. The $T$ channel, known as the null channel, is symmetric between the three interferometers and {is relatively insensitive to the GW signal.} The noise function, assumed to be equal for the $A$ and $E$ channels so that $P_{n,A} = P_{n,E} \equiv P_{n}$, is given by~\cite{Cornish:2001bb,Cornish:2018dyw}~\footnote{See also the LISA Data Challenge Manual~\cite{LISA-LCST-SGS-MAN-001}.}
\be
P_{n}(f) = \frac{1}{3}\left[2+\cos\left(f/f_0\right)\right]P_{\rm IMS}(f) + \frac{4}{3}\left[1+\cos\left(f/f_0\right)+\cos^2\left(f/f_0\right)\right] P_{\rm Acc}(f) \,,
\ee
where $f_0=1/(2\pi L)=0.019\,{\rm Hz}$ and the interferometer measurement system noise is
\be 
P_{\rm IMS}(f) = 3.6\times 10^{-41}\, {\rm Hz}^{-1} \left[1+\left(\frac{2{\rm mHz}}{f}\right)^4\right] \,,
\ee
and the acceleration noise is
\be 
P_{\rm Acc}(f) = 9.2\times 10^{-52}\, {\rm Hz}^{-1} \left(\frac{f}{{\rm Hz}}\right)^{-4} \left[1+\left(\frac{0.4{\rm mHz}}{f}\right)^2\right] \left[1+\left(\frac{f}{8{\rm mHz}}\right)^4\right] \,.
\ee
The $\mathcal{Q}_O^{ab}(\vec{k})$ matrices for the $O = A,E,T$ channels are linear combinations of those in the Michelson basis ($i = 1,2,3$): $\mathcal{Q}_O^{ab}(\vec{k}) = \sum_{i=1,2,3} c_O^i \mathcal{Q}_i^{ab}(\vec{k})$~\cite{Bartolo:2018qqn}, where
\be
c = \left(
\begin{array}{ccc}
 \frac{2}{3} & -\frac{1}{3} & -\frac{1}{3} \\
 0 & -\frac{1}{\sqrt{3}} & \frac{1}{\sqrt{3}} \\
 \frac{1}{3} & \frac{1}{3} & \frac{1}{3} \\
\end{array}
\right)
\ee
and
\be
\mathcal{Q}_i^{ab}(\vec{k}) = \frac{1}{4} e^{-i\vec{k}\cdot\vec{x}_i} \left[ \mathcal{T}(kL,\hat{k}\cdot\hat{l}_i) \hat{l}_i^a \hat{l}_i^b - \mathcal{T}(kL,-\hat{k}\cdot\hat{l}_{i+2}) \hat{l}_{i+2}^a \hat{l}_{i+2}^b \right] \,.
\ee
Here $\hat{l_i} = (\vec{x}_{i+1} - \vec{x}_i)/L$ is the unit vector pointing from spacecraft $i$ to spacecraft $i+1$, all indices $i,i+1,\dots$ are modulo $3$, and the detector transfer function is given by
\be
\mathcal{T}(kL,\hat{k}\cdot\hat{l}) = e^{-i kL (1+\hat{k}\cdot\hat{l})/2} \,{\rm sinc}\!\left[ \frac{kL}{2} (1-\hat{k}\cdot\hat{l})\right] + e^{i\pi kL (1-\hat{k}\cdot\hat{l})/2} \,{\rm sinc}\!\left[\frac{kL}{2} (1+\hat{k}\cdot\hat{l})\right] \,.
\ee
Using these we can calculate the monopole and dipole functions for LISA. 

\begin{figure}
\centering
\includegraphics[height=0.35\textwidth]{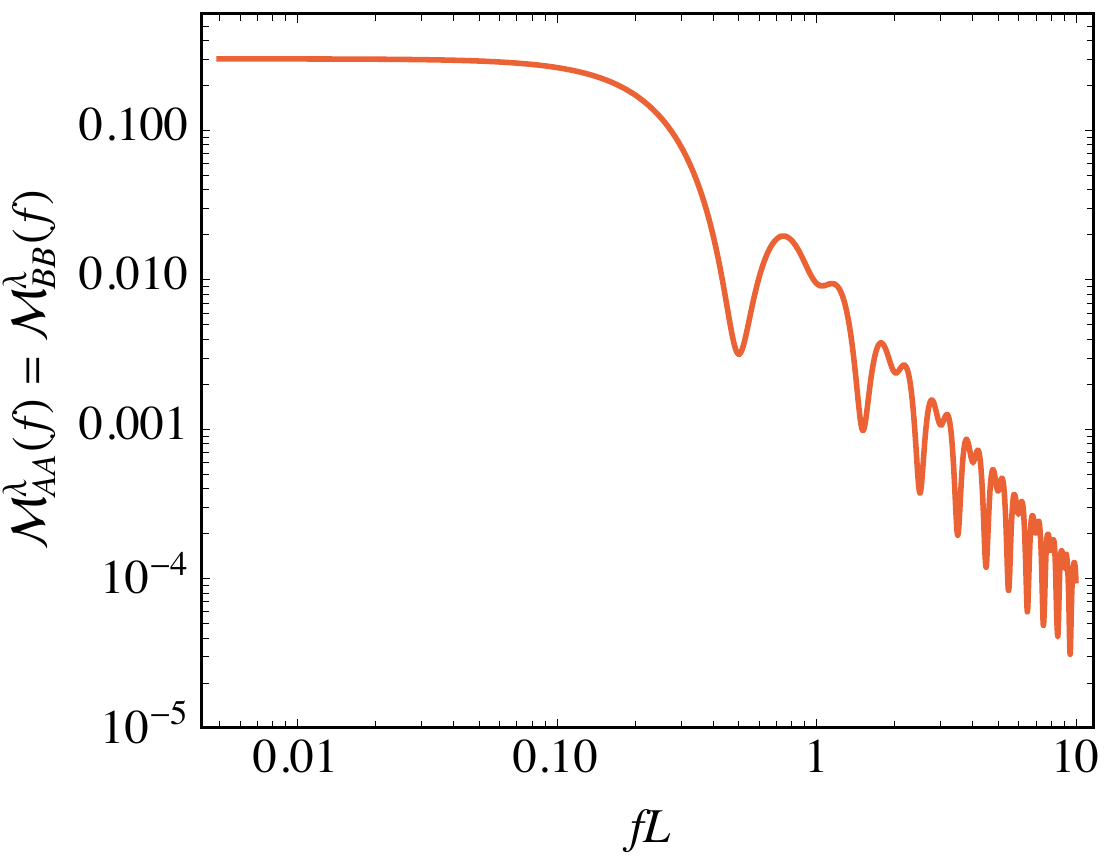} \hspace{6mm}
\includegraphics[height=0.35\textwidth]{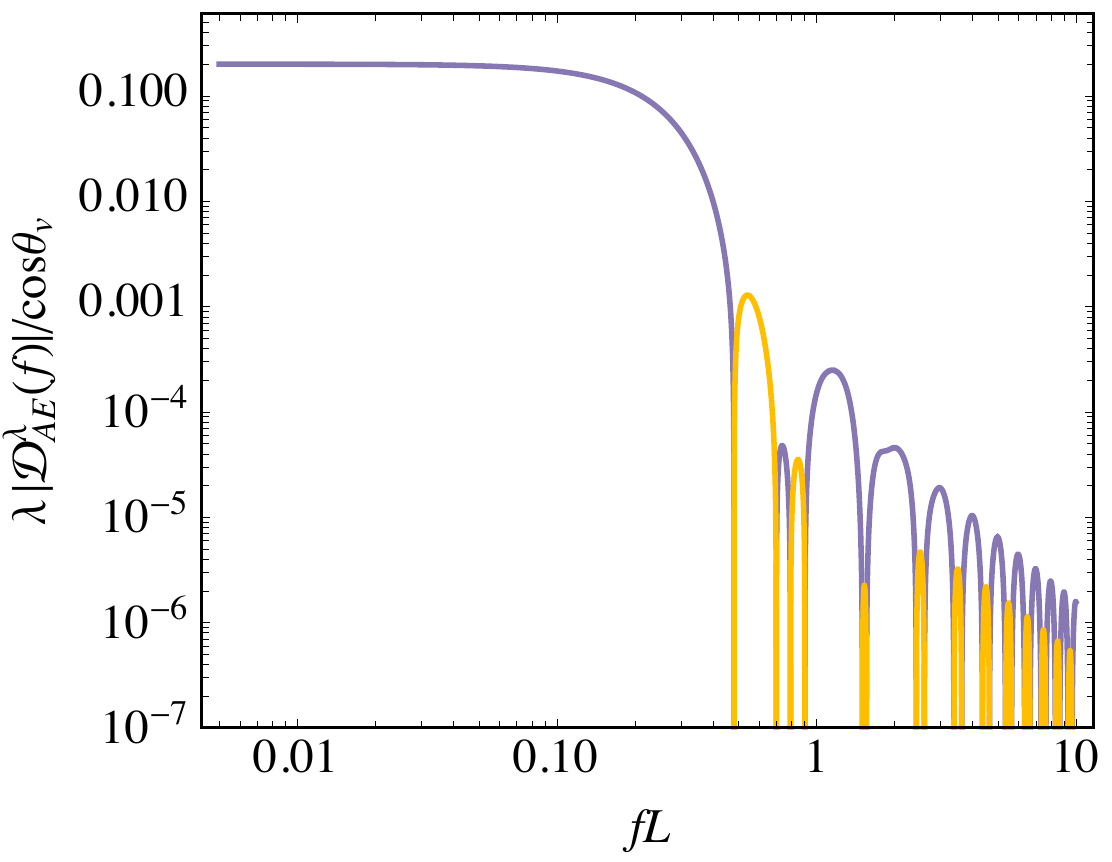}
\caption{\it The non-zero monopole response function (left panel) and the dipole response function (right panel) for LISA. In the right panel {the purple and yellow colours} indicate positive and negative values for $\lambda \mathcal{D}_{AE}^\lambda$.}
\label{fig:responsefunctions}
\end{figure}

The monopole functions $\mathcal{M}_{AA}^\lambda$ and $\mathcal{M}_{EE}^\lambda$ are equal, non-zero and independent of the helicity $\lambda$, whereas the corresponding dipole functions vanish: $\mathcal{D}_{AA}^\lambda = \mathcal{D}_{EE}^\lambda = 0$. Moreover, $\mathcal{M}_{AE}^\lambda = 0$, so the signal-to-noise ratio for LISA observing a GW monopole signal is therefore
\be \label{SNR_TOT}
{\rm SNR}_{\rm tot} = \sqrt{{\rm SNR}_{AA}^2 + {\rm SNR}_{EE}^2} = \sqrt{2}\,{\rm SNR}_{AA} \,.
\ee
The dipole response function for the $AE$ channel combination is instead non-zero. It depends on the angle $\theta_v$ between the normal of the LISA's detector plane and the direction of motion of the solar system, and the helicity of the signal, $\mathcal{D}_{AE}^\lambda \propto \lambda \cos\theta_v$. The $AE$ channel therefore probes the circular polarisation of the GW signal and the signal-to-noise ratio for observing a circularly-polarised signal with LISA is given by
\be \label{SNR_POL}
{\rm SNR}_{\rm pol} = \sqrt{{\rm SNR}_{AE}^2 + {\rm SNR}_{EA}^2} = \sqrt{2}\,{\rm SNR}_{AE} \,.
\ee
The non-zero response functions $\mathcal{M}_{AA}^\lambda = \mathcal{M}_{EE}^\lambda$ and $\mathcal{D}_{AE}^\lambda$ are displayed in Fig.~\ref{fig:responsefunctions}.

\begin{figure}
\centering
\includegraphics[height=0.35\textwidth]{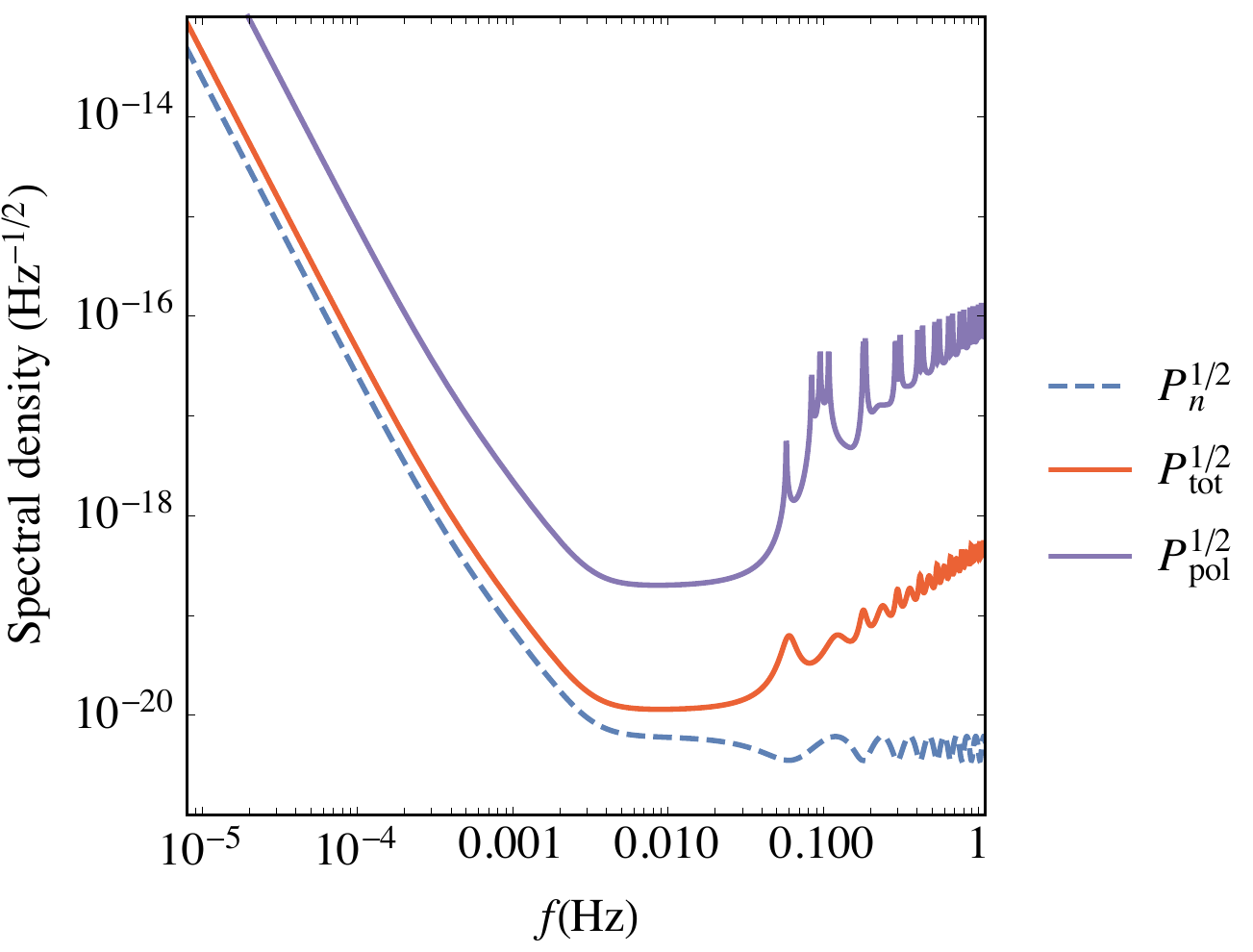} \hspace{1mm}
\includegraphics[height=0.35\textwidth]{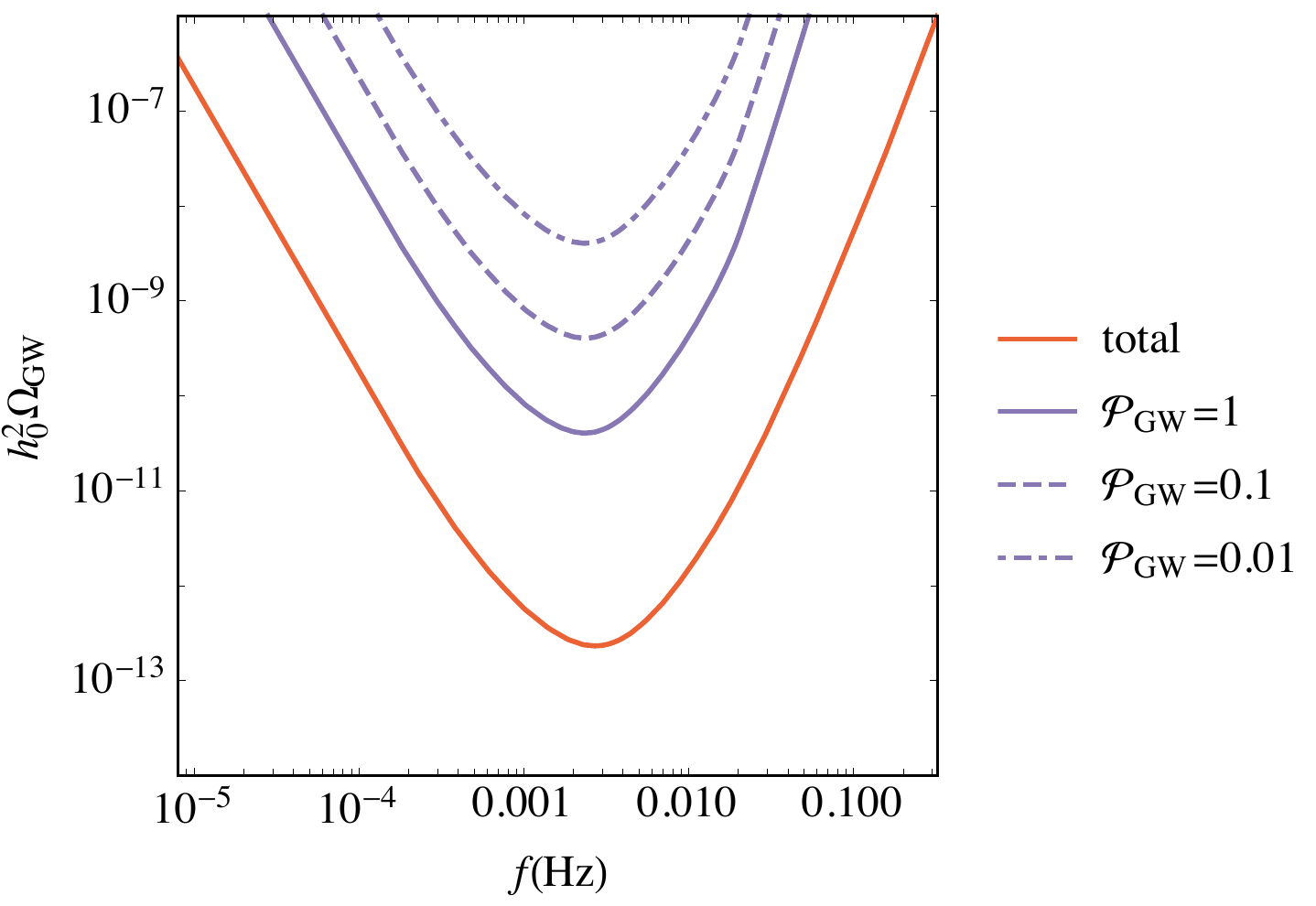}
\caption{{{\it Left panel: The dashed line shows the noise spectral density $P_{n}^{1/2}$ for LISA, the solid red line the corresponding  sensitivity for the total GW signal and the solid purple line the corresponding sensitivity for a fully polarised GW signal, $\mathcal{P}_{\rm GW}=1$. Right panel: The power-law integrated LISA sensitivity curves for the total GW signal and {polarisation of the GW signal} {for a 4-year integration time}  with different values of the polarisation fraction $\mathcal{P}_{\rm GW}$} ({\it see Eq.}~\eqref{eq:PGW}).}}
\label{fig:PI}
\end{figure}

{In the left panel of Fig.~\ref{fig:PI} we show the LISA sensitivities for the total {GW signal and its polarisation}  defined as 
\be \label{eq:sensitivities}
P_{\rm tot}(f) = \frac{P_n(f)}{\mathcal{M}_{AA}^\lambda(f)} \,, \qquad P_{\rm pol}(f) = \frac{P_n(f)}{4v\mathcal{D}_{AE}^\lambda(f)\mathcal{P}_{\rm GW}} \,,
\ee
where $\mathcal{P}_{\rm GW}$ denotes the fractional polarisation of the GW signal (see Eq.~\eqref{eq:PGW}). In the right panel of Fig.~\ref{fig:PI} we show the power-law integrated sensitivity curves for LISA assuming $t_{\rm obs} = 4\,{\rm y}$ and the threshold $\rm SNR = 10$, for different values of the polarisation fraction $\mathcal{P}_{\rm GW}$. We find that LISA can observe the polarisation of a fully-polarised GW signal down to $h_0^2\Omega_{\rm GW} = 4\times 10^{-11}$, in agreement with Ref.~\cite{Domcke:2019zls}. The sensitivity scales as a function of the polarisation fraction of the signal as $1/\mathcal{P}_{\rm GW}$.}

It is clear from the above discussion that detectors with just a single interferometer channel, such as a single LIGO detector, cannot detect circular polarisation of the SGWB. Nor, indeed, can AEDGE or a pair of such detectors. On the other hand, missions with one or more triangular sets of interferometers such as ALIA, BBO~\cite{Crowder:2005nr}, DECIGO~\cite{Kawamura:2018esd} and AMIGO~\cite{Ni:2019nau} would be sensitive to polarisation of the SGWB at higher frequencies than LISA.


\section{Results for the LISA sensitivity to the SGWB and its polarisation}
\label{sec:results}

{We now assess the sensitivity of LISA to both the SGWB and to its circular polarisation, as would be generated by helical turbulence following a first-order phase transition. To do this we take an approach with the least possible sensitivity to the underlying particle physics model, calculating the  transition strengths and temperatures, bubble sizes, and helicity fractions required to obtain a LISA SNR value greater than or equal to 10. This requires us to draw upon much of the analysis in the previous sections, first calculating the contributions to the total GW spectrum (Eq.~\eqref{Omega_GW}) from both direct cascade (Eq.~\eqref{H_stage1}) and inverse cascade (Eq.~\eqref{H_stage2}) turbulence for a particular point in parameter space and then computing the associated circularly-polarised spectrum (Eq.~\ref{pol_spectra}). Finally, we translate both signal types into the LISA SNR values associated with the total GW spectra (Eq.~\eqref{SNR_TOT}) and its polarised counterpart (Eq.~\eqref{SNR_POL}).}

\begin{figure}
\centering
    \begin{tabular}{c c}
     \vspace{0.5cm}
        \underline{\large{Total signal}} & \underline{\large{Polarised signal}} \\
        \includegraphics[width=.49\linewidth]{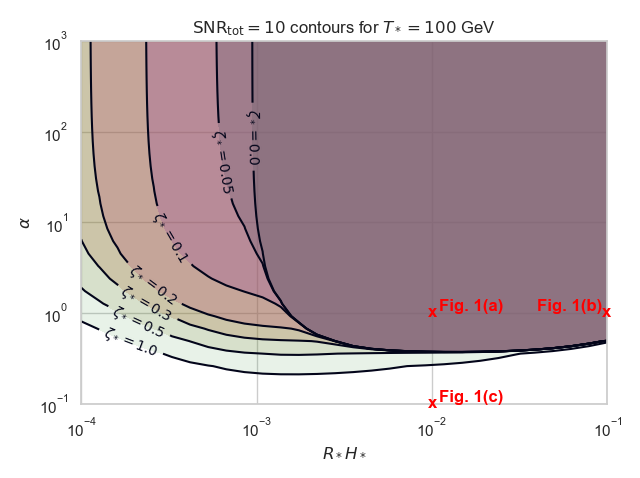} & \includegraphics[width=.49\linewidth]{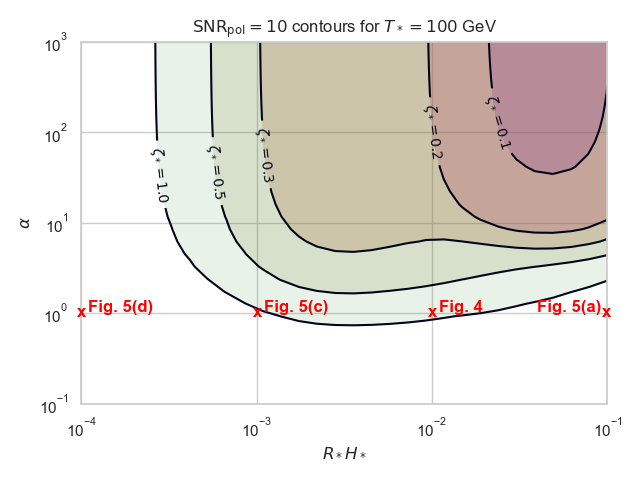} \\
        \includegraphics[width=.49\linewidth]{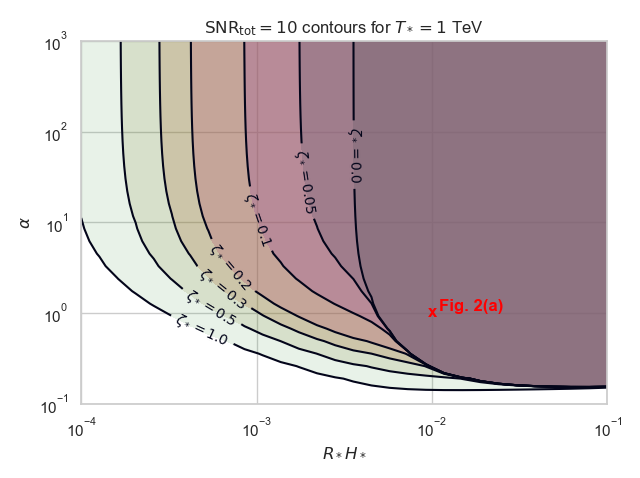} & \includegraphics[width=.49\linewidth]{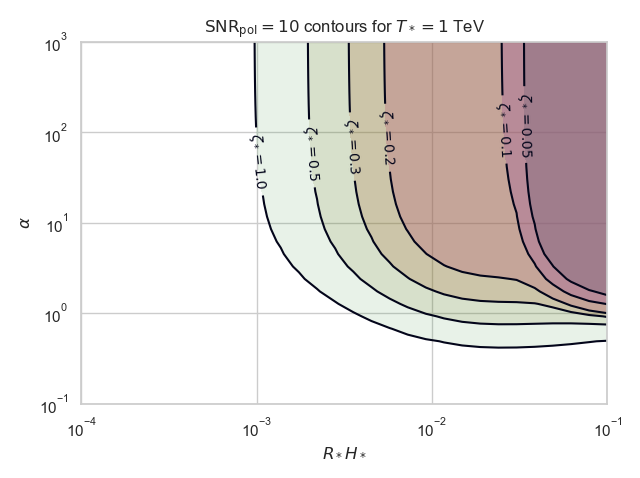} \\
        \includegraphics[width=.49\linewidth]{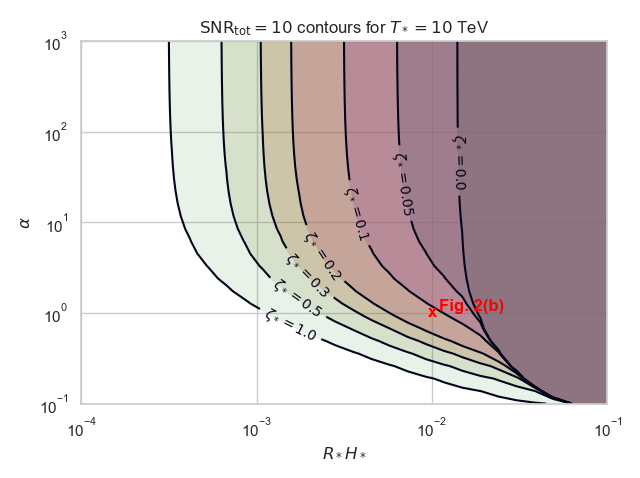} & \includegraphics[width=.49\linewidth]{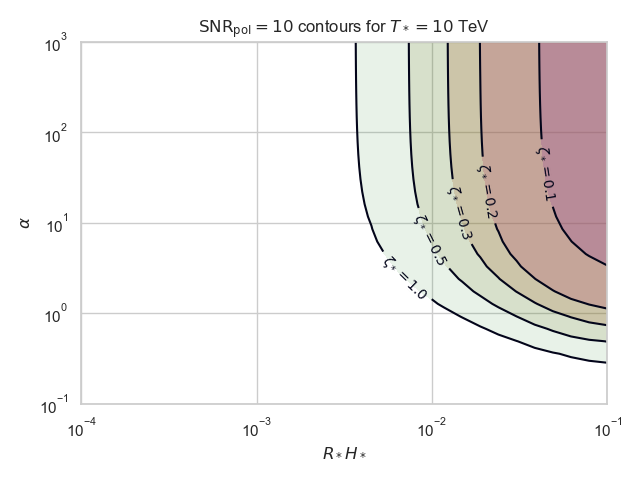}
    \end{tabular}
\caption{\it{Signal-to-noise (SNR) $= 10$ contours in the $(R_* H_*, \alpha)$ plane for $T_*=100$~GeV (top), $T_*=1$~TeV (middle) and $T_*=10$~TeV (bottom), for a LISA measurement with a 4-year observation time. In the left panels the SNR is shown for the total SGWB signal and in the right panels for observing the polarisation of the SGWB. The different contours correspond to various values of the initial helicity fraction $\zeta_*$, as shown in the plots. {The red crosses correspond to sample GW spectra for fixed $\alpha$, $R_*H_*$ and $T_*$ plotted in the indicated previous Figures, which allow comparison of the LISA sensitivity to the GW signal for ranges of $\zeta_*$ values.}}}
\label{fig:SNR}
\end{figure}

The left panels of Fig.~\ref{fig:SNR} display reaches in the $(R_*H_*, \alpha)$ plane for a LISA measurement of the overall strength of the total SGWB signal with a signal-to-noise ratio ${\rm SNR}_{\rm tot} = 10$, with larger values of SNR found {in the shaded regions} above these lines. The panels from top to bottom correspond to $T_* = 10$\,GeV, 1\,TeV and 10\,TeV, and the various contours correspond to different values of the initial helicity fraction $\zeta_*$. {Thus, if a parameter point is enclosed within the shaded area of a given $\zeta_*$ contour, one can infer that a GW signal with this value of $\zeta_*$ would be detectable by LISA with an $\textrm{SNR} \geq 10$.}  {The red crosses correspond to sample frequency spectra plotted in the indicated previous Figures for fixed $R_*H_*$, $\alpha$ and $T_*$, which exemplify the spectral sensitivity of LISA to the GW signal.} 

As expected, the largest detectable signals in all three panels come from larger values of $R_*H_*$, where the number of bubbles per horizon is smaller and thus the average bubble radius at collision is larger. As $R_*H_*$ decreases we see that increasingly large values of the helicity fraction, $\zeta_*$, are required to obtain a signal with ${\rm SNR}_{\rm tot} \gtrsim 10$. Thus GW emission from the inverse cascade turbulent period is increasingly important for LISA to be sensitive to {the total GW signal} for smaller values of $R_*H_*$. This can be traced back to the $R_*H_*$ suppression of the GW amplitude produced in a direct cascade that is a general feature of the models used to describe GW emission from turbulence (see Section~\ref{Spectrum_of_SGWB}).

We see from the different contours that increasing the initial helicity fraction increases the total SNR, in agreement with the increasing strength of the GW signal shown for different values of $\zeta_*$ {in Fig.~\ref{fig:GW_tot_spectra_fixed_temp}}. In general, the GW signal should be detectable at a level of ${\rm SNR}_{\rm tot} \gtrsim 10$ for $\alpha \gtrsim 1$ and $R_* H_* \gtrsim 10^{-3}$ for a transition at $T_*=100\,$GeV. For larger values of $T_*$, larger values of $R_* H_*$ are needed for ${\rm SNR}_{\rm tot} = 10$ measurements, though smaller values of $\alpha$ are sufficient.

We also see in the left panels of Fig.~\ref{fig:SNR} that for large values of $R_*H_*$ the position of the ${\rm SNR}_{\rm tot} = 10$ contours are approximately independent of $\zeta_*$ and depend only on $\alpha$, whereas for smaller values of $R_*H_*$ the contour lines have a greater dependence on the value of $\zeta_*$ associated with the contour. This is explained by the fact that for large $R_*H_*$ the contribution of inverse cascade period is minimised as the large average bubble radius at collision means inverse cascade turbulence cannot operate for very long before being washed out by the Hubble expansion. Thus increasing $\zeta_*$ does relatively little to increase the amplitude of the signal ({see Fig.~\hyperref[fig:GW_tot_spectra_fixed_temp]{1(b)}}) and has minimal effect on its sensitivity to LISA. Conversely, for small $R_*H_*$ the inverse cascade can operate for far longer before being washed out by the expansion, and thus the potential contribution from the inverse cascade to the total GW signal can be much larger ({see Fig.~\hyperref[fig:GW_tot_spectra_fixed_temp]{1(a)}}). As larger values of $\zeta_*$ result in greater importance of the inverse cascade period for the total GW signal, it follows that in the low-$R_*H_*$ region of the parameter space, the $\rm SNR_{\rm tot}$ is much more sensitive to the value of $\zeta_*$~\footnote{We note that new simulations suggest that the $f^1$ plateau at low frequencies could also develop in cases with small initial helicity~\cite{Pol:2019yex}. This would improve detection prospects for low-$\zeta_*$ scenarios and reduce the dependence of the total SNR on that parameter.}.

The right panels of Fig.~\ref{fig:SNR} display the corresponding reaches in the $(R_* H_*, \alpha)$ plane for a LISA measurement of the circular polarisation of the SGWB with ${\rm SNR}_{\rm pol} = 10$. As expected the reach is smaller than for the total GW signal, but we see that many of the qualitative features of the plots of ${\rm SNR}_{\rm tot}$ described above are also present in the polarised case. Whilst detection prospects for circular polarisation in the SGWB are strongest for large $R_*H_*$ where the suppression in the amplitude of the spectra is minimised (see Fig.~\ref{fig:pol_vary-RH}), in order for LISA to be able to probe a circularly-polarised signal at small $R_*H_*$, larger values of $\zeta_*$ are required to compensate for the suppression $R_*H_*$ introduces into the total GW signal. Larger $\zeta_*$ means fully-polarised GWs from the inverse cascade period make an increasingly important contribution to the total GW signal, raising the amplitude of $\Omega^{+}_{\rm GW}$ relative to $\Omega^{-}_{\rm GW}$ and increasing the prospects for detection by LISA of circular polarisation in the SGWB from a phase transition.

Comparing the right panels of Fig.~\ref{fig:SNR}, we see that for smaller values of the helicity fraction, $\zeta_* \lesssim 0.2$, LISA is most sensitive to {polarisation of the GW signal} when the transition temperature is $T_* = 1 \, \rm TeV$. This can be understood by {looking at Fig.~\ref{fig:GW_tot_spectra_vary_temp}} and noting that, of the three transition temperatures, the $T_* = 1 \, \rm TeV$ spectrum peaks at the optimal frequency to be sensitive to the fully-polarised low-frequency inverse cascade plateau that develops in the signal for small $\zeta_*$.

As seen in the right hand panels of Fig.~\ref{fig:SNR}, in the polarised case the positions of the ${\rm SNR}_{\rm pol} = 10$ contours exhibit a larger relative dependence on $\zeta_*$ at large $R_*H_*$ than their ${\rm SNR}_{\rm tot}$ counterparts. Whilst in the $\Omega^{\rm tot}_{\rm GW}$ case the low-frequency inverse cascade contribution was less important for larger $R_*H_*$, in the polarised case it has a larger impact. Even when considering the case of a low-frequency plateau in the signal associated with relatively low helicity inverse cascade turbulence (see {see Fig.~\hyperref[fig:pol]{4(a)}}), fully-polarised GW are still being emitted, implying that relatively small changes in the value of $\zeta_*$ can have a larger effect on the net polarisation of the SGWB and the ability of LISA to probe it.

As seen in the top right panel of Fig.~\ref{fig:SNR}, for larger values of the helicity fraction, $\zeta_* \gtrsim 0.3$, the parameter space in which ${\rm SNR}_{\rm pol} > 10$ expands significantly in the $T_*=100 \rm GeV$ case, allowing a larger range of small $R_*H_*$ values to be probed by LISA. This can be understood by referring to the $T_* = 100 \rm GeV$ spectrum plot {in Fig.~\hyperref[fig:GW_tot_spectra_fixed_temp]{1(a)}} and noting that for intermediate values of the helicity fraction, $0.1 \lesssim \zeta_* \lesssim 0.5$ the low-frequency, fully-polarised inverse cascade plateau transitions into a new, distinct peak, and in so doing becomes rapidly more sensitive to the frequency band where LISA is most sensitive.

Similar behaviour is seen for the $T_* = 1 \, \rm TeV$ case shown in the middle panel of Fig.~\ref{fig:SNR}, though less pronounced, because the GW spectra for this transition temperature peak at higher frequencies where LISA is already more sensitive to the fully-polarised inverse cascade plateau.

\section{Conclusions}
\label{sec:conx}

We have analysed in this paper the prospects for detection of circular polarisation of the stochastic GW background produced by a first-order phase transition at a temperature $T_* \ge 100$~GeV. We focused on an analytical model for the sourcing of GWs by MHD turbulence produced in the plasma during the transition. Crucially, the model allows us to describe not only a direct energy cascade in the plasma but also an inverse cascade that develops if there is some initial helicity fraction in the fluid motion, usually sourced from helicity left over in the magnetic field after the transition. The direct cascade describes energy transferred into smaller scales, and the resulting signal peak corresponds to the characteristic scale of the transition, which is related to the average bubble size $R_*$.

If some initial helicity fraction is present in the plasma, then the helical component of the energy in the MHD turbulence will be conserved during the direct cascade, whilst the non-helical part will be dissipated away at small scales due to the plasma's intrinsic viscosity. This results in a period of inverse cascade MHD turbulence following the direct cascade, where the turbulence is fully helical and energy in the turbulence is instead transferred to increasingly large scales. This process lasts for the rest of the Hubble time, and continuously produces a GW signal forming a plateau in wavelength that extends to the horizon size. Critically, the signal produced during this second stage of the turbulence will be circularly polarised and, provided the initial helicity fraction is large enough, it results in an overall stochastic background with a significant degree of polarisation.

We have compared our results with the more common description of fluid dynamics involving a sound wave period followed by a top-hat approximation to the turbulence. The crucial difference between the models is the characteristic scale, which depends in the model we use not only on the bubble size but also the Mach number. This induces a dependence on the average fluid velocity which becomes lower in weaker transitions, causing the GW spectrum to peak at lower frequencies.

We have revisited the capability of future GW detectors to measure the polarisation of a SGWB background, focusing on LISA. We find that in the model we use to describe the signal arising from MHD turbulence following a phase transition, detection with ${\rm SNR}_{\rm pol} > 10$ could be possible. However, this would require a sufficiently strong phase transition with $\alpha \gtrapprox 1$ as well as either a large helicity fraction close to unity or large bubbles of sizes approaching the horizon size. The smaller the helicity fraction, the more supercooled the transition would have to be to produce an observable polarisation in the stochastic GW background signal.

We conclude that LISA may have a significant opportunity to measure polarisation of the SGWB. However, we emphasise several caveats. The strength of any such signal is sensitive to the strength of the underlying first-order phase transition and the sizes of the bubbles it produces. In particular, potentially observable signals would require a significant amount of supercooling, potentially leading to new difficulties in modeling the turbulence not yet taken into account. Moreover, the chances of such a measurement depend crucially on the seeding of some helical turbulence in the primordial plasma. We emphasise also that the model we have used to calculate the amount of polarisation certainly requires improvement, and should be tensioned against other models as they emerge. We look forward to improvements in understanding any possible cosmological first-order phase transition and the possible origin and magnitude of helical turbulence, and improvements in modelling their consequences.

\section*{Acknowledgements}

The work of JE, MF, ML and VV was supported by the United Kingdom STFC Grant ST/P000258/1. Also, JE received support from the Estonian Research Council grant MOBTT5, ML was partly supported by the Polish National Science Center grant 2018/31/D/ST2/02048,
and would also like to acknowledge hospitality and support from KITP at UCSB,
where part of this work was carried out, supported in part by the National Science Foundation under Grant No. NSF PHY-1748958. MF and AW were funded by the European Research Council under the European Union's Horizon 2020 programme (ERC Grant Agreement no.648680 DARKHORIZONS), and VV was partly supported by the Estonian Research Council grant PRG803.
ML, VV and AW would also like to express gratitude to the organizers of the  {\it Gravitational Waves from the Early Universe} programme at NORDITA for providing a great working environment from which this project benefited, as well as partial support during the programme.

\bibliographystyle{JHEP}
\bibliography{Mag}

\end{document}